\documentclass[sigplan, nonacm]{acmart}
\renewcommand\footnotetextcopyrightpermission[1]{}
\usepackage{graphicx} % Required for inserting images
\usepackage{tikz}
\usepackage{xcolor}
\usepackage{cleveref}
\usepackage[labelfont=small,textfont=small]{subcaption}
\usepackage{booktabs}
\usepackage{multirow}
\usepackage{tabularx}
\usepackage{paralist}
\usepackage{enumitem}
\usepackage[noframe]{showframe}
\usepackage{url}
\usepackage{array}
\usepackage{listings}
\usepackage{fancyvrb}
\usepackage{texments}
\usepackage{ragged2e}
\usepackage{xspace}

\newif\ifcomments
\commentstrue
\ifcomments
    \providecommand{\swang}[1]{{\protect\color{purple}{\bf [swang: #1]}}}
    \providecommand{\megan}[1]{{\protect\color{orange}{\bf [megan: #1]}}}
\else
    \providecommand{\swang}[1]{}
    \providecommand{\weixin}[1]{}
    \providecommand{\megan}[1]{}
\fi

\newcommand{\sys}{Piper\xspace}

\newcommand{\chunk}{\texttt{Chunk}\xspace}
\newcommand{\comm}{\texttt{Comm}\xspace}

\newcommand{\chunks}{\texttt{Chunk}s\xspace}
\newcommand{\comms}{\texttt{Comm}s\xspace}

\newcommand{\shard}{\texttt{Shard}\xspace}

\newcommand{\order}{\texttt{Order}\xspace}

\usepackage{listings}
\usepackage{xcolor}

\definecolor{codebg}{RGB}{248,248,248}
\definecolor{codekw}{RGB}{0,0,180}
\definecolor{codestr}{RGB}{163,21,21}
\definecolor{codecom}{RGB}{0,128,0}
\definecolor{codenum}{RGB}{120,120,120}

\lstdefinestyle{pyglobal}{
  language=Python,
  backgroundcolor=\color{codebg},
  basicstyle=\ttfamily\scriptsize,
  keywordstyle=\color{codekw}\bfseries,
  stringstyle=\color{codestr},
  commentstyle=\color{codecom}\itshape,
  numbers=left,
  numberstyle=\tiny\color{codenum},
  numbersep=8pt,
  stepnumber=1,
  showstringspaces=false,
  tabsize=2,
  breaklines=true,
  breakatwhitespace=true,
  frame=single,
  framerule=0.5pt,
  rulecolor=\color{black!30},
  captionpos=b,
  xleftmargin=1em,
  xrightmargin=0.5em,
  aboveskip=0.8em,
  belowskip=0.8em,
  keepspaces=true
}

\usepackage{pifont} % for \ding
\newcommand{\cmark}{\ding{51}}
\newcommand{\xmark}{\ding{55}}
\newcommand{\astmark}{\ensuremath{\ast}}

\usepackage{algorithm}
\usepackage{algpseudocode}

\usepackage{caption}
\title{\sys{}: A Programmable Distributed Training System}
\author{Megan Frisella}
\affiliation{%
  \institution{University of Washington}
  \city{Seattle}
  \state{WA}
  \country{USA}
}

\author{Shubham Tiwari}
\affiliation{%
  \institution{University of Washington}
  \city{Seattle}
  \state{WA}
  \country{USA}
}

\author{Andy Ruan}
\affiliation{%
  \institution{University of Washington}
  \city{Seattle}
  \state{WA}
  \country{USA}
}
\author{Yi Pan}
\affiliation{%
  \institution{University of Washington and Shanghai Jiao Tong University}
  \city{Seattle}
  \state{WA}
  \country{USA}
}

\author{Parker Gustafson}
\affiliation{%
  \institution{University of Washington}
  \city{Seattle}
  \state{WA}
  \country{USA}
}

\author{Mat Jacob}
\affiliation{%
  \institution{University of Washington}
  \city{Seattle}
  \state{WA}
  \country{USA}
}

\author{Gilbert Bernstein}
\affiliation{%
  \institution{University of Washington}
  \city{Seattle}
  \state{WA}
  \country{USA}
}

\author{Stephanie Wang}
\affiliation{%
  \institution{University of Washington}
  \city{Seattle}
  \state{WA}
  \country{USA}
}

\begin{document}

\begin{abstract}
% As machine learning (ML) model architectures have increased in size, ML training now requires scaling across hundreds to thousands of accelerators. This is challenging because there are many ways to shard and replicate the model, and each introduces different and possibly interacting communication overheads. There is no one-size fits-all solution, as the right strategy depends on the workload and hardware. This problem requires choosing a high-level parallelism strategy then scheduling and coordinating each device’s low-level execution plan. These challenges must be solved jointly to maximize overall compute efficiency and hide communication overheads. Our goal is to allow practitioners to easily realize arbitrary high- and low-level execution plans, but in a way that is flexible and adaptable to new workloads. Our key insight is to decoupling what the overall execution strategy should be from how it is realized by the runtime. We propose a unified representation of communication and computation in a training DAG, a compiler for lowering this to a low-level execution plan with user hints and a runtime capable of executing arbitrary training schedules. We show that \sys{} matches the performance of general-purpose frameworks on widely-supported training schedules and optimizations, while improving performance and memory efficiency for specific workloads by tuning optimizations for composed parallelism strategies.

Large-scale model training increasingly relies on composing multiple parallelism strategies, such as data, pipeline, and expert parallelism, together with memory-saving optimizations like ZeRO.
Deployed systems for foundation model pretraining often rely on human experts to manually design a high-level parallelism strategy then implement the corresponding low-level execution strategy, making it difficult to adapt the system to new strategies.
Meanwhile, many general-purpose frameworks are more flexible but their implementations are still tied to a fixed set of common parallelism strategies, making it challenging to integrate state-of-the-art strategies.
% However, existing systems either hard-code a narrow set of high-performance strategies or expose interfaces whose implementations are tightly coupled to specific parallelism dimensions, making it hard to express and optimize new strategies.

We present \sys{}, a user-controllable distributed training system that decouples the strategy from the runtime implementation. 
\sys{} allows users to declare a comprehensive distributed training strategy with a small set of model annotations and scheduling directives.
Each directive applies a transformation on \sys{}'s intermediate representation (IR), a unified global training DAG that represents all computation and communication.
Using this IR, \sys{} compiles per-device execution plans and executes them with a distributed runtime agnostic to the strategy. 
We show that the combined system maintains performance parity on commonly available strategies such as ZeRO, while also enabling additional performance and memory efficiency gains through joint scheduling of compute and communication in composed parallelism strategies such as DeepSeek-V3's DualPipe.
% \sys{}'s IR enables throughput and memory-efficient combinations of PP with ZeRO memory optimizations and EP, while maintaining performance parity on commonly supported schedules.
% \sys{} supports X-X\% larger batch sizes for PP x DP compared with TorchTitan and for PP x EP improves 13\% over its own default schedule and 6-30\% over baseline systems.
\end{abstract}

\maketitle
\pagestyle{plain}

\vspace{-10pt}

\section{Introduction}

As machine learning (ML) models have increased in size, pretraining now requires scaling across hundreds to thousands of accelerators.
This is challenging because there are many ways to shard and replicate the model, and each introduces different and possibly interacting communication overheads.
For example, modern workloads now use combinations of data (DP), tensor (TP), expert (EP), context (CP) and pipeline (PP) parallelism together with memory-saving optimizations such as ZeRO~\cite{zero-deepspeed}. There is no one-size fits-all solution, as the right strategy depends on the workload and hardware.

A workload's distributed training strategy can be decomposed into: (1) the high-level parallelism strategy, such as the dimensions along which to shard and replicate the model and activations, and (2) each device's low-level strategy for executing the compute and communication operations dictated by the high-level parallelism strategy.
The former strategy determines the minimum memory, compute, and communication load per-device, which in turn upper-bounds the overall system throughput.
The latter strategy determines how close the system can get to the upper bound by effectively scheduling each device's resources.
% For example, each device should overlap compute and communication operations, allocate additional memory to enable pre-fetching when feasible, and reduce stalling across devices.
The two strategies together determine the workload's actual throughput.

The sheer size of the combined space of strategies has so far made full automation intractable.
Thus, while some solutions can find an optimal strategy within a particular subspace~\cite{chen2024centauri, peng2019generic}, the systems deployed in practice still rely overwhelmingly on human experts.
For example, DeepSeek-V3 introduced DualPipe, a custom PP schedule that when composed with EP enables each device to use local micro-batch overlapping to hide EP communication overheads~\cite{deepseekv3-deepseekai}.
This solution required human-engineered codesign of the high-level parallelism strategy with a hand-implemented per-device execution strategy to manage intra-GPU resources, such as the streaming multiprocessors (SMs) allocated to compute vs.~communication.
Such systems require high effort; experts must design \emph{and implement} a fixed strategy that is specialized to a particular model and cluster.
This results in a system that is hard to adapt to new strategies.

On the other hand, general-purpose frameworks such as Megatron~\cite{megatron-nvidia}, DeepSpeed~\cite{deepspeed-microsoft}, and TorchTitan~\cite{liang2025torchtitan} offer a more flexible and model-agnostic interface, with knobs for tuning the distributed training strategy.
However, these frameworks eagerly dispatch operations for each high-level parallelism dimension as if the dimensions are independent, making it challenging to jointly schedule operations from composed strategies.
For example, conceptually DualPipe shares a GPU between two PP microbatches; this is challenging to implement in existing frameworks that assume that each microbatch is allocated the full GPU.
% They also do not correctly combine PP with arbitrary ZeRO levels, which can result in unexpected out-of-memory errors.
While compiler-based frameworks such as JAX/XLA present a more generic tensor placement abstraction~\cite{gspmd-google} instead of a fixed set of knobs, they cannot easily support arbitrary PP schedules nor control over each device's resources.
Thus, again it requires high effort to introduce novel strategies such as DualPipe.

The key problem we address is extensibility in distributed training systems.
Our goal is to build a system that minimizes the effort needed to specify and implement an arbitrary distributed training strategy.
Thus, we build a system that provides good performance for common strategies and control when needed for higher performance, making it amenable to both expert engineering and automated search.
Our key insight is to decouple \emph{what} the overall execution strategy should be from \emph{how} it is realized by the runtime.

The challenges are to: (1) design a user scheduling API that enables simple specification yet sufficient control over the distributed training strategy, (2) design an intermediate representation (IR) to interface between the user API and system runtime, and (3) build an efficient system runtime to execute an arbitrary strategy.

% We introduce a unified abstraction for specifying the strategy.
Inspired by previous works on SPMD-style tensor annotations~\cite{gspmd-google} and PP scheduling~\cite{pippy2022,jaxpp-nvidia,piper}, we first propose an API for placing different tensors in an arbitrary PyTorch model.
This can be used to specify high-level parallelism strategies such as the pipeline-parallel stage boundaries or whether to use expert parallelism.

Next, we design an API for specifying the corresponding low-level execution strategy for the execution order of all operators on each device.
A strawman solution is to ask the user to fully specify the strategy by writing their own scheduler for operators over intra-GPU resources such as CUDA streams.
This provides users with full control but is impractical.
On the other hand, the API must also be expressive enough to capture novel strategies for intra-GPU resource allocation such as DualPipe.

% that captures semantically meaningful model chunks (e.g., pipeline-parallel stages and expert-MLP layers).
% From this specification, we first compile a global training DAG that captures the required data dependencies the model segments.
% Next, the user specifies the high-level strategy, which dictates the placement of model chunks, and a partial low-level execution strategy, which dictates the granularity, resource assignment, and execution order of compute and communication operations.
% These low-level decisions can have significant impact on the overall run time, but the choice can be challenging to make automatically.
% For example, in DualPipe, background all-reduces from DP can add 46\% interference to EP communications; partitioning into smaller all-reduces reduces interference but in other cases could increase run time due to lower communication efficiency.
% Such patterns are difficult for the system to accurately model ahead of time.
% On the other hand, asking the user to fully specify the low-level execution strategy is not practical.

To address this challenge, we expose a set of directives for lowering the system's IR to a per-device execution strategy.
We design an IR that captures all compute and communication operations in a global training DAG.
The compiler begins by extracting a non-distributed DAG from the model code.
Then, each directive specified by the user transforms the IR.
Directives include transforms such as splitting batches into microbatches to increase overlap opportunity and assigning resources such as device streams to chunks.
To provide the user more control when needed, we allow the user to specify ordering constraints between compute nodes.
We then use a generic scheduling policy to decide the execution order on each device.
This enables the user to control the granularity of scheduling and make some decisions while the systems fills in the rest.
We guarantee the safety of transformations, i.e. each user directive should be compatible with the original high-level strategy.
% We guarantee that the lowering does not break the original high-level strategy by guarding for “safe” transformations, e.g. transformations should not affect parameter sharding or PP microbatch ordering.

Finally, we build an efficient and flexible distributed runtime to execute the combined strategy.
We use a centralized scheduler to produce and distribute each device's local execution strategy.
Each worker loads its associated model weights, then performs local scheduling to allocate shared device resources such as memory, communicators, and GPU streams.
Compared to the original model definition, the resulting execution of tensor operators may be distributed, overlapped, out-of-order, etc.;
\sys{} guarantees that the global execution plan still respects the data and temporal dependencies specified by the model definition and user directives.

% We implement our approach in the \sys{} compiler and runtime. The compiler consumes a TorchDynamo compute graph and user annotations / lowering flags, producing a per-device execution plan. The runtime loads  \sys{} matches the performance of baseline training frameworks on widely-supported training schedules and optimizations, while also enabling intra-device bucket- and microbatch-overlapping.

Thus, we present \sys{}.
We implement \sys{} as a \linebreak torch.compile~\cite{pytorch2} backend and use Ray to implement the distributed runtime.
We show that \sys{} can match the performance of generic training frameworks on widely supported strategies, while further enabling concise expression of novel strategies such as DualPipe, producing an overall 6-30\% improvement in throughput across baselines including Megatron and TorchTitan.
We also demonstrate that \sys{} can better support arbitrary parallelism strategies through a case study combining PP with different ZeRO levels.
While other generic training frameworks fail to support certain combinations, \sys{} can support all strategies, enabling 3-8x larger batch sizes.
Our contributions include

\begin{itemize}
    \item A user scheduling interface for controlling the high- and low-level execution strategy for inter- and intra-device parallelism. %, including the placement, granularity, resource assignment, and ordering of operators. 
    \item A unified IR for global training DAGs that enables joint scheduling of communication and computation inserted by different parallelism strategies. 
    \item An efficient distributed runtime that is strategy-agnostic.
    % \item System-level support for lowering a high-level training schedule to a low-level execution plan via user directives. The system fills in gaps to output an absolute ordering, overlapping and resource allocation of operators by exploiting overlapping opportunities while avoiding resource contention. 
    \item Our end-to-end evaluation shows that \sys{} matches the performance of general-purpose frameworks on widely supported strategies, while improving performance and memory efficiency on composed strategies. 
\end{itemize}

\begin{figure}[t]
    \centering
    \includegraphics[keepaspectratio,width=0.95\linewidth,trim=4.25cm 4.25cm 4.25cm 4.25cm, clip,page=1]{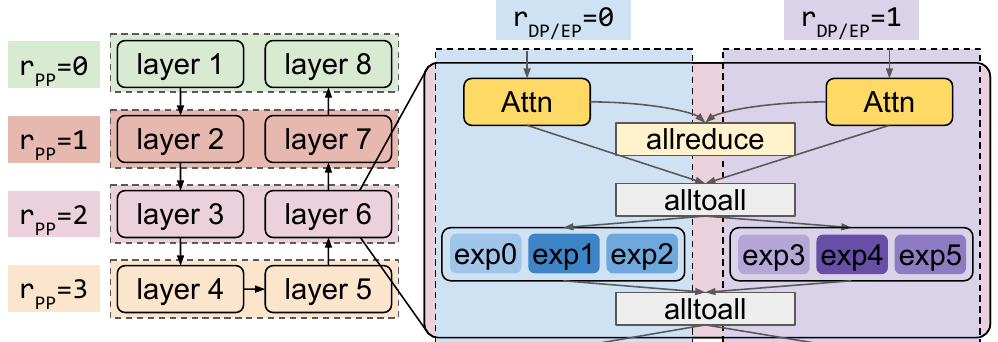}
    \caption{DualPipe example combining PP-4 across layers with EP-2 for the expert layer and DP-2 for the non-expert attention layer. This placement uses PP stage interleaving; PP rank 0 gets the first and last layers, etc.
    We show a variant of DualPipe known as DualPipeV~\cite{qi2025dual} that places each layer on one PP rank instead of two.}
    \label{fig:parallelism}
\end{figure}

\section{Background}

% \subsection{Parallelism strategies}

We first overview some possible choices in high-level parallelism strategy.

\paragraph{DP and ZeRO} In DP~\cite{tensorflow,ddp-torch}, each worker holds a replica of the model weights, computes a local gradient, then averages the gradients across all replicas using an allreduce (or allgather and reduce-scatter). ZeRO~\cite{zero-deepspeed}, also known as fully sharded DP~\cite{fsdp-pytorch},  is a DP variant that reduces redundant state across workers by sharding the optimizer, gradients, and/or the model weights across DP ranks. States are rematerialized with allgather when needed, and resharded by either dropping the remote shards or reduce-scatter in the case of gradients. Communication operations can be overlapped with computation of the preceding or following compute operations.

\paragraph{TP, EP, and CP} Each worker holds a shard of either the model weights, e.g., a subset of experts in EP~\cite{lepikhin2020gshard} or a subset of rows/columns in TP~\cite{megatron-nvidia}, or the activations, e.g., a subsequence of the context in CP~\cite{context_parallelism}. Workers execute collective communication according to the sharding plan and tensor operations to ensure that all activations are correctly aggregated. Compared to DP, these collectives execute on the critical path of a batch’s computations.

\paragraph{PP} Each worker holds a shard of the model layers. Because of the sequential dependency between model layers, PP schedules use multiple microbatches to overlap execution across devices. The performance of a PP strategy depends on a complex set of factors including bubbles between devices and the amount of communication overhead. Thus, many PP schedules have been proposed to balance between such factors~\cite{gpipe, pipedream, zerobubble}. PP adds send-receive communications when activations are passed between ``stages’’, i.e. shards placed on different devices. Unlike the above parallelism strategies, PP is most easily executed with an MPMD (multiple program multiple data) approach due to each rank needing to execute a different set and order of operations.

\begin{figure}[t]
    \centering
    \includegraphics[keepaspectratio,width=0.51\textwidth,trim=1.5cm 5.2cm 1.5cm 2.5cm, clip]{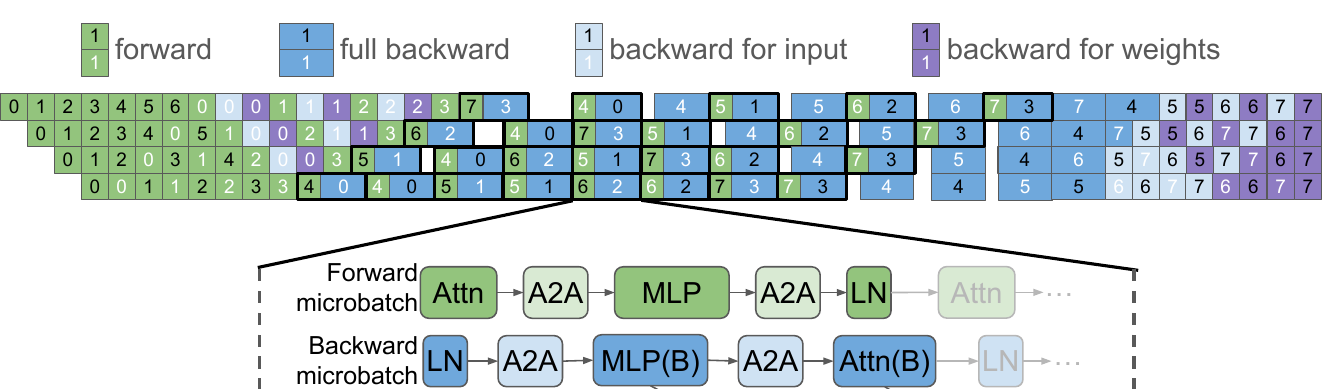}
    \caption{DualPipeV schedule~\cite{qi2025dual} for 4-way PP, 8 microbatches, 8 layers. The placement is shown in \Cref{fig:parallelism}. Combined forward-backward microbatches in bolded boxes get overlapped.}
    \label{fig:dualpipe}
\end{figure}

\begin{figure}[t]
    \centering
    \includegraphics[keepaspectratio,width=0.49\textwidth,trim=3.5cm 2.5cm 1.5cm 2.3cm, clip,page=8]{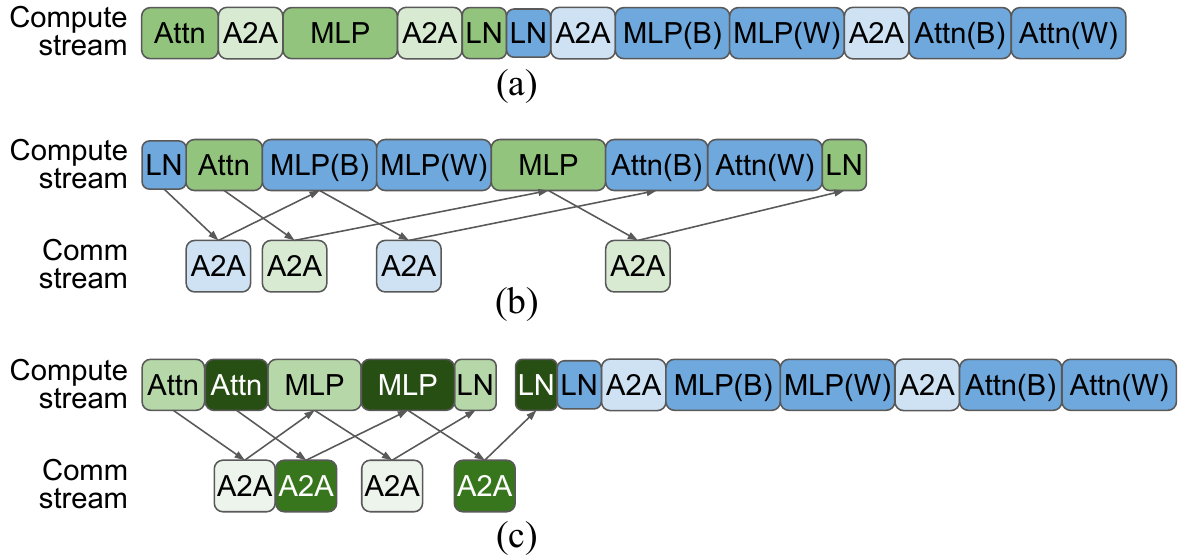}
    \caption{Different strategies for applying intra-device parallelism. \textbf{(a)} Serial forward-backward execution. \textbf{(b)} Interleaving a forward and backward microbatch can effectively hide all-to-alls.
    \textbf{(c)} Further microbatching within the forwards enables overlapping but reduces hardware efficiency.
    % \swang{could cut (c) from figure for space}
    }
    \label{fig:microbatch}
\end{figure}

\paragraph{Composition}
Given the need to scale distributed training to larger models and clusters, common practice uses multiple parallelism dimensions to ensure that models can fit in memory and maximize hardware bandwidth. For example, \Cref{fig:parallelism} shows a DualPipe-like placement that combines PP across layers with EP for the expert layer and DP for the non-expert attention layer.
The PP schedule~(\Cref{fig:dualpipe}) is carefully designed to enable overlapping of pairs of forwards and backwards microbatches~(\Cref{fig:microbatch}b).
This strategy is a variant of DualPipe known as DualPipeV~\cite{qi2025dual}; it uses microbatch overlapping as in DualPipe but places each stage on one PP rank instead of two.

As models are becoming more heterogeneous~\cite{mmllms}, the need to be able to specify different strategies per submodule is also becoming critical. For example, the Qwen3-Next models use diverse attention layers~\cite{qwen3-coder-next}, and multimodal models use modality-specific encoders/decoders composed with the transformer backbone~\cite{clip-model, llava-model, disttrain}.

% \swang{memory-saving techniques such as activation checkpointing?}

\vspace{-10pt}

\begin{figure}[t]
    \centering
    \includegraphics[keepaspectratio,width=0.45\textwidth,trim=4cm 0.8cm 4cm 1cm, clip,page=17]{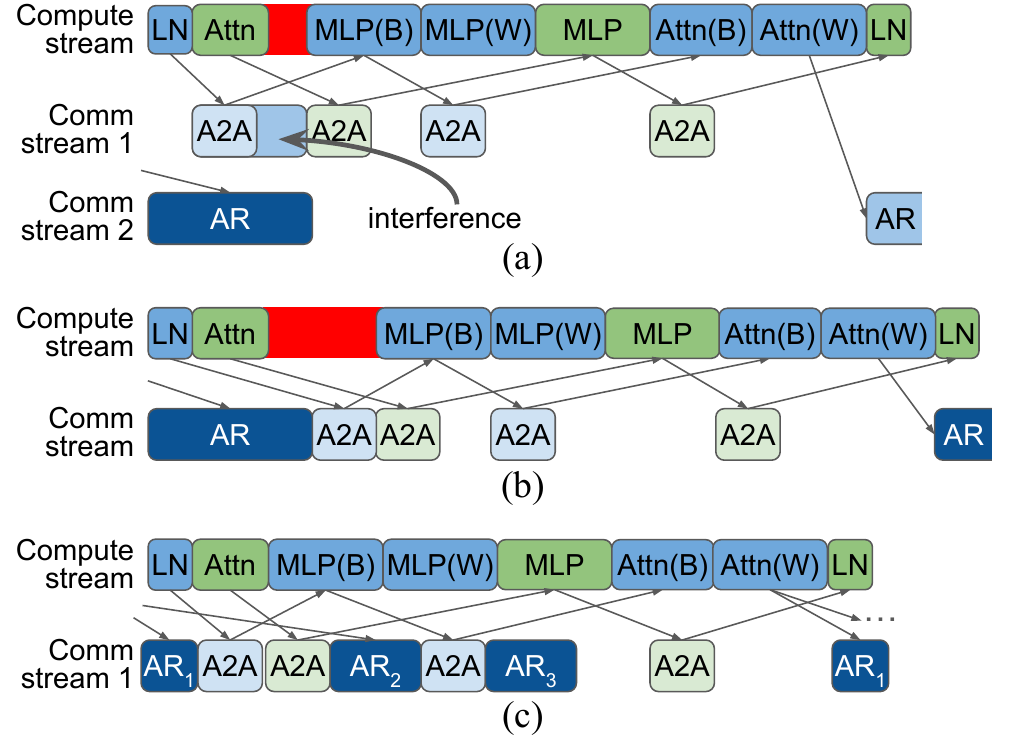}
    \caption{Different low-level execution strategies corresponding to an overlapped microbatch pair in DualPipe~(\Cref{fig:parallelism,fig:dualpipe}), now including the allreduce from using DP for Attn (dark blue AR boxes).
    \textbf{(a)} Using a separate stream for DP communication can add interference to all-to-alls.
    \textbf{(b)} Scheduling DP communication on the stream delays EP all-to-alls.
    \textbf{(c)} Partitioning DP communication into finer-grained operations and scheduling on the same stream as EP all-to-alls can produce the best run time. 
    % Slicing the allreduces into smaller operations enables scheduling them between the EP all-to-alls to avoid both interference and delaying the all-to-alls, as shown in (c).
    }
    \label{fig:interference}
\end{figure}

\section{Challenges}

\paragraph{High-level parallelism strategy}
Previous work addressing the problem of expressing high-level parallelism strategies includes GSPMD~\cite{gspmd-google} for composed tensor sharding and replication plans~(\Cref{fig:parallelism}) and complementary solutions~\cite{jaxpp-nvidia,pippy2022,piper} for the inter-device PP schedule~(\Cref{fig:dualpipe}).
We additionally observe that as communication overheads increase with larger models, the need for \emph{intra-device} parallelism has become critical.
This is commonly done by overlapping communication with compute.
This can be done across distinct microbatches, as seen in DualPipeV~(bolded boxes in \Cref{fig:dualpipe}), as well as other systems that apply local microbatching for specific operators~\cite{jangda2022breaking,wang2022overlap,chang2024flux,asynctp}.

However, the decision of when to apply intra-device parallelism is nontrivial. For example, overlapping forwards and backwards microbatches improves throughput in the DualPipeV schedule~(\Cref{fig:dualpipe}), but overlapping forwards with forwards or backwards with backwards could introduce bubbles that outweigh the gains in pairwise microbatch throughput, as shown in \Cref{fig:microbatch}c. 
% Also, aggressive microbatching enables new overlapping opportunities but can reduce hardware efficiency~\cite{xxx,dynaflow?}, as shown in \Cref{fig:microbatch}c. 
Recent work further shows that for heterogeneous models in particular, the intra-device parallelism strategy needs to be codesigned with the PP placement~\cite{tessera-osdi26}; optimal overlapping would thus require extensive profiling of the different scheduling options.

Thus, we advocate for an abstraction that allows users to flexibly specify the high-level inter- and intra-device plan, allowing both human users and automated systems to easily control the system’s behavior.

\paragraph{Low-level execution strategy.}
Once the high-level parallelism strategy has been decided, the corresponding work must be efficiently scheduled onto each accelerator's resources: compute, memory, and network. For GPU-based programming models, this requires allocating \emph{streams} and \emph{communicators} to enable concurrent execution of different kernels and communication operations, respectively.

In simple cases, such as when the model and activations fit comfortably in memory or when using only one parallelism dimension, scheduling is straightforward. 
For example, DP only adds communication off of the critical path of execution, so achieving good utilization requires only: (1) using one stream each for compute and communication, (2) choosing a communication bucket size that is large enough to ensure good network utilization but small enough to enable overlapping with compute, and (3) minimizing additional memory usage from allreduce buffers.

However, when multiple parallelism dimensions are used, each can add operations that may contend for the same resources, such as communication operations contending on network bandwidth or intermediate buffers contending for GPU memory. Resource contention can produce performance behaviors that are hard to predict.

For example, \Cref{fig:interference} shows different low-level execution strategies corresponding to an overlapped microbatch pair in DualPipe~(\Cref{fig:parallelism,fig:dualpipe}), now including the allreduce from using DP for Attn. \Cref{fig:interference}a shows the simplest and common case, which uses a separate stream for DP communication. This avoids sequential execution, which would delay the EP all-to-alls~(\Cref{fig:interference}b). However, because all-to-all and allreduce both use network resources, \Cref{fig:interference}a can also exhibit unpredictable interference.
For example, we measured a 1.46$\times$ slowdown in EP communications due to background allreduces from DP in a DualPipe strategy.
Partitioning into smaller all-reduces could reduce interference~(\Cref{fig:interference}c) but in other cases could increase overall run time due to lower communication efficiency.

% In this case, one approach to mitigate this is to slice the allreduces into smaller operations and schedule them between the EP all-to-alls~(\Cref{fig:interference}c. Although the allreduce efficiency may be reduced, we show that this can lead to an overall improvement of X\% throughput~(\Cref{sec:xxx}).

We aim to design an abstraction that allows the user to easily control this behavior, and in the future generalize to let the user control tradeoffs between interference and maximizing communication bandwidth.
The goal is to concisely capture the low-level execution strategies shown in \Cref{fig:microbatch,fig:interference}, among others.
In particular, we focus on expression of scheduling decisions on the tensor operator graph, which is compatible with lower-level optimizations such as kernel fusion that can further boost performance.

% \swang{this layer needs to be designed with the runtime, e.g., for things like stream allocations}

\paragraph{Distributed runtime.}
The distributed runtime is responsible for executing the compiled low-level execution strategy on each device.
Thus, it must: (1) remain agnostic to the strategy, (2) add minimal scheduling overhead, to avoid blocking accelerators on CPU-based scheduling, and (3) allocate the shared resources of each accelerator efficiently, without needing to ask the user to specify the exact schedule of low-level tensor operators.
% and (4) scale to thousands of accelerators.

Critically, the system should be able to jointly schedule \emph{all} of the communication operations introduced by different high-level parallelism dimensions, instead of simply scheduling each as if they were independent.
This is critical for overall throughput, especially when resources are contended.

% System performance can also be hard to predict in limited memory scenarios.
For example, we observed that when near capacity, the PyTorch memory allocator can introduce expensive device stalls waiting for in-flight work to finish so it can reclaim memory~\cite{pytorch-stalls}. Thus, memory-saving sharding techniques such as ZeRO can also significantly improve throughput, not just memory efficiency.
% \swang{we didn't directly show this in the evaluation but maybe that's okay} 
Meanwhile, we also found that existing general-purpose training frameworks do not in fact fully support all ZeRO levels when used in combination with other parallelism strategies such as PP. We believe this is due to interactions between the pre- and post-layer hooks that ZeRO introduces and PP execution, which runs each layer multiple times for one global batch. We show that by using a unified abstraction that can correctly capture and schedule these interactions with higher memory efficiency, we can support 3-8$\times$ larger batch sizes~(\Cref{sec:eval:zero}).
\section{Design}
\label{sec:design}

% To address the challenge of enabling control over execution in distributed training, our key insight is to decouple what the overall execution strategy should be from how it is realized by the runtime. 
% We propose a unified intermediate representation of a global training DAG of computation and communication operators. 
% We build a compiler which translates user models and scheduling intent into a training DAG and a runtime which correctly and efficiently executes arbitrary DAGs on a set of distributed workers. 

\begin{figure}[t]
    \centering
    \includegraphics[keepaspectratio,width=0.45\textwidth,trim=4.1cm 2.25cm 4.1cm 2.4cm, clip,page=9]{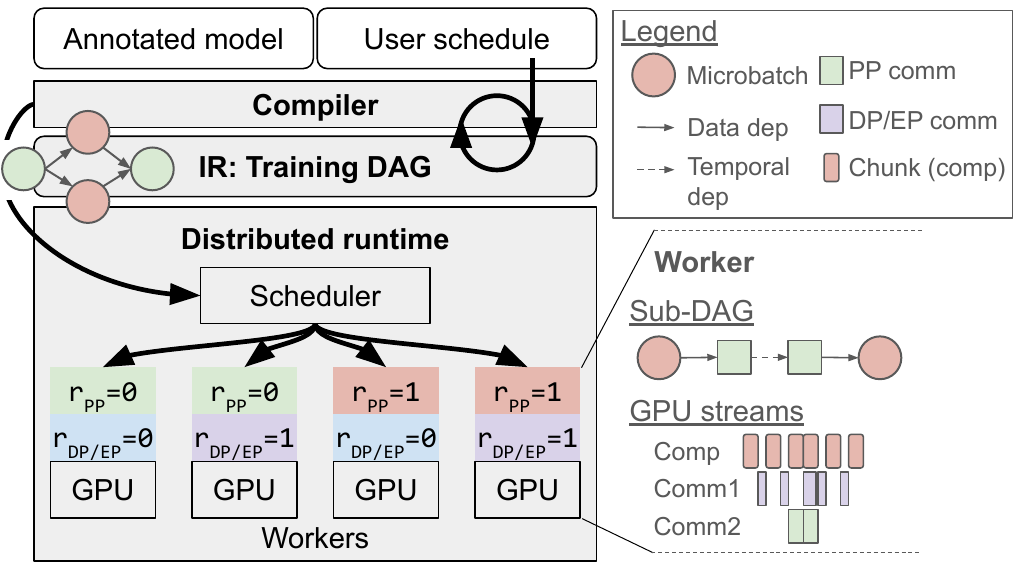}
    \caption{\sys{} overview. White rounded boxes are user inputs. Grey areas are system.}
    \label{fig:arch}
\end{figure}

\sys{} has two components~(\Cref{fig:arch}). First, the compiler translates annotated user models and user scheduling intent into a distributed execution plan. Second, the runtime executes the plan on distributed workers in a way that is agnostic to the specific execution strategy. Together, these components let \sys{} represent and realize a rich set of training schedules without hard-coding specific parallelism strategy compositions or optimizations into the runtime.

In the rest of this section, we use a simplified example based on DeepSeek-V3’s DualPipe schedule applied to an MoE transformer model.
In this example, the user combines PP=2 and EP/DP=2 and uses microbatch overlap to hide EP all-to-all latency.

\vspace{-10pt}

\subsection{API}

The API exposes the parts of execution that users may want to control, such as device placement, communication granularity, and operator ordering, while leaving the remaining low-level decisions to the system.
Each API translates to a transformation on \sys{}'s intermediate representation (IR).
The \sys{} IR is the global training DAG, a unified abstraction for distributed training that encodes computation, communication, and resource assignment, i.e. a GPU stream.

\sys{} realizes the API through two components:
\begin{compactenum}
    \item Annotations let users identify schedulable regions of model computation, which later become compute nodes in the training DAG.
    \item Scheduling directives transform the training DAG.
\end{compactenum}

\begin{figure}
    \centering
    \includegraphics[keepaspectratio,width=0.42\textwidth,trim=8cm 4.5cm 8cm 2.7cm, clip]{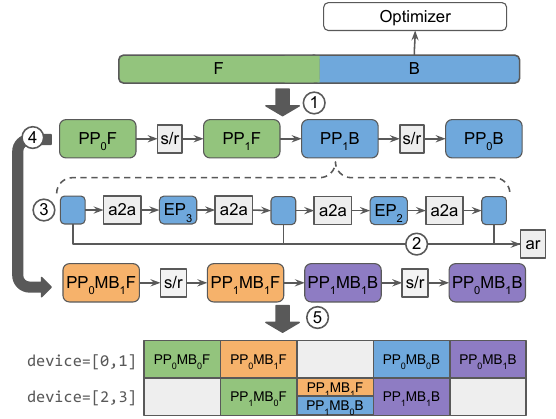}
    \caption{Transformations on the DAG IR. Each rounded box is a \chunk. Boxes with corners are \comms. 
    % Ovals are \buffers. For simplicity \buffers and the optimizer are only shown in the initial DAG. 
    Thin arrows are data dependencies. Wide arrows are transforms corresponding to scheduling directives. The schedule at the bottom shows the execution order once all data dependencies specified by the model and temporal dependencies specified by \texttt{Order} have been applied.}
    \label{fig:dag-rewrites}
\end{figure}

\paragraph{IR}
To show how \sys{}'s user scheduling APIs are implemented, we first describe the IR, the global training DAG~(\Cref{fig:dag-rewrites}).
Nodes represent coarse-grained compute or communication units and data flows along edges.
\sys{} manages memory for the data along the edges.
All communication is explicit in the graph. This allows the compiler and runtime to reason about communication, computation, and resources such as GPU memory in a unified way.

Each node in the DAG is one of the following: (1) a \chunk: the most basic unit of compute with no interleaved communication, or (2) a \comm node, which can be a point-to-point or collective operation.
%or (3) a \buffer such as a model weight parameter or gradient.
% \swang{the original text here didn't include the Buf, but the figure showing the DAG does. Should try to reconcile it.}\swang{I still have some trouble reconciling this definition of chunk with the idea of nested chunks, like a PP chunk can contain multiple DP/EP chunks. Maybe chunk could be defined as no compute for that particular parallelism dimension, e.g., a PP chunk can't contain any sends/recvs inside it.}
Nodes have a device or device mesh placement. Except for p2p \comm nodes, all nodes must have the same placement as their inputs and outputs. 

Each \chunk and \comm node has an associated \texttt{exec} function to run upon dispatch by the runtime.
This is a PyTorch \texttt{fx.Graph} of tensor operators for forwards passes, an opaque graph inserted by the PyTorch autograd engine for backwards passes, or a NCCL communication kernel inserted by \sys{}.
\chunk and \comm nodes also have an assigned logical GPU stream to run on, which the \sys{} runtime later translates to a physical GPU stream~(\Cref{sec:design:runtime}).

Conceptually, the compiler begins with a single-device DAG with a single \chunk, extracted from the model definition.
The single node has all model operators for a forwards-backwards pass as one \chunk, with no communication (the large F|B \chunk in \Cref{fig:dag-rewrites}).
The user's annotated regions of the model replace the single chunk with smaller chunks.
Each directive in the user's schedule then repeatedly transforms the DAG, until we have the lowered distributed training DAG to pass to the runtime~(\Cref{fig:arch}).

\vspace{-5pt}

\paragraph{Annotations}
\sys{}’s annotation API lets the user tag parts of the model for later identification during scheduling. An annotation identifies a semantically meaningful region of computation, such as a pipeline stage or an expert MLP block. These annotations are translated by the compiler into \chunks that can later be reordered, replicated, sharded, or otherwise transformed.

The annotation API takes in a string \texttt{dim} to allow users to add new parallelism dimensions and refer to them in scheduling directives.
For example, in Listing \ref{lst:annotations}, we create a wrapper around an MoE transformer model that annotates PP stages with the \texttt{PP} dimension tag and expert-MLP components with the \texttt{EP} dimension tag.
\sys{} infers indices for repeated annotations based on the order in the model's dataflow.
For example, the first \texttt{PP} annotation gets index 0 and the second \texttt{PP} annotation gets index 1.
% For example, the \texttt{EP} annotation is repeated for every \texttt{MoELayer} so each the \texttt{EP} block in layer 1 gets \texttt{EP}=0, and so on.

\begin{lstlisting}[float, caption={MoE model annotated with PP and EP components.}, label={lst:annotations}]
PP = "pp_tag"; EP = "ep_tag"
class TransformerModel:
    def forward(self, x) ->
        with sys.annotate(PP):
            h = self.embeddings(x)
            h = self.layer2(self.layer1(x))
        with sys.annotate(PP):    
            h = self.layer4(self.layer3(x))
            h = self.output(h)
        return h

class MoELayer:
    def forward(self, x):
        x = self.router(x)
        with sys.annotate(EP):
           x = self.experts(x)
        return x
\end{lstlisting}

% \begin{figure}
%     \centering
%     \includegraphics[keepaspectratio,width=0.49\textwidth,trim=2cm 3cm 2cm 3cm, clip,page=1]{figures/chunk-dag.pdf}
%     \caption{Training DAG Listing \ref{lst:annotations} model distributed according to the Listing \ref{lst:user-schedule} schedule. \swang{TODO}}
%     \label{fig:dag}
% \end{figure}

\vspace{-11pt}

\paragraph{Scheduling directives }
\sys{} exposes a small set of scheduling directives for controlling placement, communication granularity, and operator overlapping and ordering.
We use filters to allow users to apply directives to one or multiple \chunks at a time.
A filter can include zero to multiple dimension names, plus a value to filter on for the dimension index.
For example, \texttt{(PP=0)} refers to the first PP stage and \texttt{(EP=3)} refers to the \texttt{MoELayer} in layer 3.
``$*$'' pattern-matches on all cases and ``$-$'' excludes all cases of a tag. 
For example, \texttt{(PP=1, EP=-)} matches on all non-expert components of PP stage 1 in Listing \ref{lst:annotations}.
For shorthand, omitting a tag from the filter entirely will match on all occurrences of that tag. 
% For example, \texttt{(PP=*, EP=-)} and \texttt{[(EP=-)]} are the same filter. 

By default, the system also supports a dimension named \texttt{PASS}, to indicate a specific part of the forwards-backwards pass for the same \chunk.
Valid values are \texttt{F} for forwards, \texttt{B} for full backwards, \texttt{Bi} for backwards for inputs and \texttt{Bw} for backwards for weights.
% \megan{unclear how Bw/Bi get introduced}
This is useful for cases like ZeroBubble~\cite{zerobubble}, a PP schedule that splits the backwards pass into two stages for better pipelining across devices.

% \begin{table}[t]
% \centering
% \begin{tabular}{l c}
% \hline
% Strategy & Rewrites \\
% \hline
% PP        & Place, Split, Order \\
% DP        & Replicate \\
% EP        & Shard \\
% \hline
% \end{tabular}
% \caption{Directives for expressing different parallelism strategies. DualPipe~\cite{deepseekv3-deepseekai} uses all of these directives.}
% \label{tab:parallelism}
% \end{table}

The placement directives are \texttt{Place}, \texttt{Replicate}, and \texttt{Shard}, as described next.
Placement directives take in a device mesh (an N-D array of devices) which represents devices with unique integers.
% Devices are currently specified by unique integers.\swang{todo for future is to be able to specify the device mesh and how it maps to the actual GPU topology.}
Placement filters must have \texttt{PASS=*} (included by default), i.e. we cannot have different placements for the forwards and backwards passes of the same \chunk.

Placement directives may add communication nodes to the training DAG.
Thus, we also allow these directives to pass a logical GPU stream that any inserted communication operations should execute on.
During execution, the runtime will create one physical stream per logical stream on the participating devices.
If no stream was provided, then the system uses the default stream, the same stream on which the compute operators run.
This allows users to have full control over stream execution without having to manage streams themselves.

Next we describe each directive in terms of how it transforms the training DAG~(\Cref{fig:dag-rewrites}).

\texttt{\textbf{Place}(filters, devices, stream=None)}.
A placement directive that updates the device placement of matched nodes.
If multiple devices are passed, then additional placement directives are needed to determine the concrete device mapping for each \chunk.
If a matched node is adjacent to another node that has a different device placement, add a P2P send/recv \comm~((1) in \Cref{fig:dag-rewrites}) assigned to \texttt{stream}. 

\texttt{\textbf{Replicate}(filter, devices, gather\_stream=None, reduce\_stream=None, shard\_params=False, \\shard\_grads=False, bucket\_sz=None)}.
A placement directive that replicates the matched nodes across devices.
Insert a collective \comm after the backward pass (or backward for weights pass) for each matched \chunk to synchronize gradients among the replicas of the \chunk~((2) in \Cref{fig:dag-rewrites}). 
The operation is all-reduce by default, and alternatively reduce-scatter when \texttt{shard\_grads=True}.
When \texttt{shard\_params=True}, insert an all-gather \comm before every matched node.
New gather \comms are assigned to \texttt{gather\_stream} and new reduce \comms are assigned to \texttt{reduce\_stream} for fine-grained control over overlappable \comms.

% When no \texttt{bucket\_size} is provided, the default behavior is to launch the communication operations for all matched \chunks as one coarse-grained bucket.
% In the DAG IR, this is done by adding a data dependency edge from each matched \chunk's backwards pass to a single communication operation (or vice versa for all-gather).
% Thus, applying \texttt{Replicate} to the entire model would launch an allreduce for all model weights at once, effectively disabling overlapping.
The user can optionally pass in a \texttt{bucket\_size}, and the \sys{} runtime will attempt to bucket communications to not exceed this size.
If the bucket size is smaller than the weights for a \chunk, the system transforms the DAG by breaking each \chunk into smaller \chunks.
If the user wants more control over buckets, they can use multiple \texttt{Replicate} rules.
For example, to ensure that each model layer is launched as one bucket, the user could annotate each layer of the model with \texttt{dim=layer}, then run \texttt{Replicate} on each \texttt{layer} index.

\texttt{\textbf{Shard}(filter, devices, stream=None)}.
A placement directive that shards the weights associated with matched \chunks along dimension 0.
Currently, this requires that the preceding or subsequent \chunk has the same \texttt{devices} but with the \texttt{Replicate} rule.
Inserts an all-to-all \comm before and after each matched \chunk and updates upstream/downstream data dependencies to flow through the new \comms~((3) in \Cref{fig:dag-rewrites}). 
This directive combined with \texttt{Replicate} can be used to represent EP with DP/ZeRO.

In future work, we will generalize our approach and infer communication for arbitrary combinations of these directives, e.g., by incorporating GSPMD-style sharding~\cite{gspmd-google}. 

\texttt{\textbf{Split}(filter, dim, num\_microbatches)}. 
Replicate the matched nodes \texttt{num\_microbatches} many times~((4) in \Cref{fig:dag-rewrites}).
This adds a new dimension with the user-provided name.
For example, Listing \ref{lst:user-schedule} shows an example where the user specifies 2 PP microbatches under the dimension name \texttt{MB}.
\texttt{MB}=0 and \texttt{MB}=1 refer to the first and second microbatches, respectively. 
This rewrite requires that the filtered nodes form a contiguous sub-DAG.

\texttt{\textbf{Order}(filter\_list)}. 
Add a dependency between each pair of adjacent filters.
For example, if two adjacent filters match sub-DAGs $A$ and $B$ respectively, then this adds a temporal edge from the topologically last node(s) in $A$ and the topologically first node(s) in $B$ to enforce an ordering between the two sub-DAGs.
This requires that each filter matches a contiguous sub-DAG.
This directive can be useful for expressing arbitrary PP schedules as well as other ordering dependencies.

To express overlapped execution of different microbatches, the user can pass in a nested list of filters.
For example, in Listing \ref{lst:user-schedule}, line 11 specifies that the forwards for microbatch 0 of PP stage 1 should execute before an overlapped pair comprising the forwards for microbatch 1 of PP stage 1 and the backwards for microbatch 0 of PP stage 1.
In this case, the \sys{} runtime will interleave the two sub-DAGs of matched \chunks and \comms~(\Cref{sec:design:runtime}).

Specifying PP schedules manually can be burdensome.
Thus, in practice, we expect that users would use functions that build the orders for a given PP schedule. %, set of PP stage annotations on the model, and microbatch tags. 

% \begin{lstlisting}[caption={A simplified user schedule for DualPipe.}, label={lst:user-schedule}]
\begin{lstlisting}[
  caption={A simplified user schedule for DualPipe.},
  label={lst:user-schedule},
  numbers=left,
  numberstyle=\scriptsize,
  breaklines=true,
  breakatwhitespace=true,
  columns=fullflexible,
  keepspaces=true
]
PPStr, EPStr, DPStr = sys.stream(), sys.stream(), sys.stream()
Place((PP=0), device=[0,2], stream=PPStr)
Place((PP=1), device=[1,3], stream=PPStr)
Replicate((PP=0, EP=-), devices=[0,2], reduce_stream=DPStr)
Replicate((PP=1, EP=-), devices=[1,3], reduce_stream=DPStr)
Shard((PP=0, EP=*), devices=[0,2], stream=EPStr)
Shard((PP=1, EP=*), devices=[1,3], stream=EPStr)
MB = "microbatch"
Split((), dim=MB, num_microbatches=2)
Order([(PP=0, MB=0, PASS=F), (PP=0, MB=1, PASS=F),                (PP=0, MB=0, PASS=B), (PP=0, MB=1, PASS=B)])
Order([(PP=1, MB=0, PASS=F), [(PP=1, MB=1, PASS=F),               (PP=1, MB=0, PASS=B)], (PP=1, MB=1, PASS=B)])
\end{lstlisting}

Next, we put the directives together to walk through the example user program in Listing \ref{lst:user-schedule}, which shows a simplified DualPipe-like schedule~\cite{qi2025dual}.
Line 1 creates one logical stream each for PP, DP, and EP communication.
Lines 2-3 place the \chunks for pipeline stages 0 and 1 on different devices, applying the transformation in (1) in \Cref{fig:dag-rewrites} to insert P2P operators at cross-device boundaries.
% Lines 4-7 replicate and shard the model to apply DP and EP.
Lines 4-5 apply DP to all non-expert weights, across devices $[0,2]$ for pipeline stage 0 and devices $[1,3]$ for pipeline stage 1.
This appends an allreduce per PP stage to the backwards \chunks, as shown in (2) in \Cref{fig:dag-rewrites}. 
Lines 6-7 shard the experts over the same devices as the DP replicas, which inserts an all-to-all collective before and after each expert chunk, as shown in (3) in \Cref{fig:dag-rewrites}.

Line 9 splits the model into two microbatches ((4) in \Cref{fig:dag-rewrites}). 
\texttt{Split} duplicates the entire DAG because its filter matches everything, assigning each DAG copy a unique index within the new \texttt{MB} dimension.
Line 10 assigns the microbatch ordering depicted in (5) in \Cref{fig:dag-rewrites}. 
Line 11 designates that $PP_1MB_1F$ and $PP_1MB_0B$ are overlappable by passing them together in a nested list. 

The schedule also expresses intent about resource assignment. Communication originating from the PP rewrite gets its own \texttt{PPStream}, meaning that send/recv operations between PP stages are overlappable with compute. DP and EP also use separate streams, so they also will not synchronize with each other.
This will yield the execution plan shown in \Cref{fig:interference}a.
Using the same stream for DP and EP would yield \Cref{fig:interference}b.
\Cref{fig:interference}c could be supported by using the same stream for DP and EP and a smaller communication bucket size for \texttt{Replicate}.
Currently, the \sys{} scheduler may still schedule allreduces even though they would delay the critical path.
In the future, we plan to additionally support more control via \texttt{Order} for specific communication operations and/or smarter system scheduling. 
% In the future, we plan to also support ordering constraints between communication operations to express the strategy shown in \Cref{fig:interference}c; this would simply require a way to reference specific communication operations. 

\vspace{-5pt}

\subsection{\sys{} Compiler}
\label{sec:design:compiler}

The \sys{} compiler translates the user's annotated model and scheduling directives into the DAG IR.
The compiled DAG specifies a partial ordering of \chunks and \comms that includes data dependencies from the model dataflow and temporal dependencies from the \texttt{Order} directive.
Each \chunk and \comm is also assigned a stream but not yet an order in the stream.
The goal is to produce a plan that can be mechanically executed by the distributed runtime with simple heuristics~(\Cref{sec:design:runtime}).

Compilation proceeds in two phases. First, \sys{} extracts user-annotated model regions as coarse-grained \chunks and builds an initial single-device training DAG. Second, it applies the user’s scheduling directives as graph rewrites to produce a distributed training DAG with explicit communication and finer-grained execution. 
%Third, \sys{} fills in remaining ordering or resource assignments to take advantage of resource overlapping. \megan{moving this to the runtime section}

\vspace{-5pt}

\paragraph{Phase 1: Model annotations and chunk extraction}
\label{sec:chunk-extraction}
\sys{} first extracts the dataflow graph of tensor operators.
Currently this is done using TorchDynamo~\cite{pytorch2}, a PyTorch JIT compiler which uses symbolic tracing to build \texttt{fx.Graph}s made up of PyTorch tensor operators.
% Piper assumes the operator graphs do not include communication (a local model implementation). 
Initially, all tensor operators are put into one forwards-backwards \chunk~(the large F|B chunk in \Cref{fig:dag-rewrites}).

% \begin{itemize}
%     \item \textbf{Input}: Annotated model
%     \item \textbf{Output}: Single-device DAG of annotated compute chunks
% \end{itemize}

Next, \sys{} splits the full-model operator graph at annotation boundaries~(Listing~\ref{lst:annotations}) to produce an operator subgraph per finer-grained chunk. 
This sub-\texttt{fx.Graph} serves as a forward \chunk's \texttt{exec} function, and the PyTorch autograd engine implicitly adds the executable graph for backwards \chunks.
\sys{} encodes forward \chunk dependencies according to the data dependencies extracted from the model definition. 
Backward \chunk dependencies follow in reverse order.

% Starting from an annotated model, \sys{} extracts annotated compute regions as \chunks. The resulting initial training DAG is single-device: it captures the computation and dependency structure implied by the model and the annotations, without introducing distribution-specific communication or resource placement.

The PyTorch tensor operator graph also determines model weight dependencies for each \chunk.
\sys{} uses this to associate a \textit{bucket} of model state (parameters, gradients, and optimizer state) with each \chunk.
After this initial pass, the compiler will have produced a single-device training DAG split into the user-annotated \chunks.

\paragraph{Phase 2: DAG transformations}
Next, \sys{} mechanically applies the DAG rewrites as directed by the user schedule~(Listing~\ref{lst:user-schedule}).
% If a directive's preconditions are not met, then \sys{} throws an error to the user.
% For example, \texttt{Split} requires that the given filters match to a contiguous sub-DAG.
% 
\sys{} may insert new \comm nodes during the rewrite process; \sys{} assigns the \texttt{exec} function for \comms to the communication kernel to execute, e.g., all-to-all or allreduce.

% \sys{} mechanically applies DAG rewrites as directed by the user schedule.
% Each directive performs an in-place modification of graph nodes and/or edges. 
% For rewrites which have requirements about what patterns they can match, \sys{} checks that the conditions are satisfied. 

Currently, we require that a model state bucket can only be associated with \chunks that have the same placement.
Associating a single bucket with \chunks that have different placements (e.g., tied embeddings) could be supported by inserting additional gradient synchronization between the \chunks' backward passes.

After applying user transforms, the compiler applies a pass to elide unnecessary communication from parameter bucket rematerialization.
In particular, the \texttt{Replicate} directive naively inserts an allgather before each \chunk that consumes the data.
In cases where two consecutive \chunks use the same weights, \sys{} collapses these into one allgather.
Similarly, for ZeRO-2 gradient sharding, if two consecutive \chunks accumulate to the same gradient bucket, then \sys{} collapses the reduce-scatter into one operation.
These elisions do not add any memory overhead and reduce unnecessary communication time.

The final output of this pass is a global training DAG that reflects communication and inter- and intra-device placement.
It captures data dependencies from the model definition and temporal dependencies introduced by \texttt{Order}.
Each \chunk has a \texttt{device} and \texttt{stream} assignment.
\sys{} validates that all device assignments are present; future work could automatically propagate from adjacent nodes to fill in missing assignments, similar to GSPMD~\cite{gspmd-google}.
Missing stream assignments are assigned to the default compute stream.

\subsection{\sys{} Runtime}
\label{sec:design:runtime}

\subsubsection{Centralized scheduler}

The \sys{} runtime executes arbitrary training DAGs on a set of distributed workers. 
Execution starts at the centralized scheduler, which takes in the transformed training DAG that provides a partial ordering and resource assignment (device and stream) for \chunks and \comms.
The goal is to produce a partial ordering per device.
The ordering is partial because \chunks and \comms on the same stream are totally ordered, while \chunks and \comms on different streams are only ordered if they have a data or temporal dependency.

First, the scheduler decomposes the training DAG into one unique sub-DAG per PP rank.
Workers with the same PP rank execute with SPMD so they receive the same sub-DAG.
For example with PP-2 x EP-2, we use 4 workers and 2 unique sub-DAGs~(\Cref{fig:arch}).

% This decomposition distinguishes between MPMD-style directives like Shard(mesh=[0]) and SPMD-style directives such as Shard(mesh=[[0,1]]) and Replicate. Replicate and Shard duplicate or partition portions of the graph across worker groups, while Place assigns different parts of the graph to different workers. Because of this distinction, inter-device micro-batching can lead to heterogeneous patterns across pipeline ranks even when other dimensions execute in an SPMD style.

Decomposing the DAG simply involves gathering the sub-DAG of nodes for each device.
Compute operators that execute on different devices are connected by a \texttt{send-recv} node; this decomposes into a \texttt{send} for the sending rank and a \texttt{recv} for the receiving rank.
Ranks with the same PP stage replicate all communication operations within that stage.
% : the \texttt{recv} or \texttt{send} before or after the stage and any collective communication operations within the stage.

The sub-DAG includes only the data dependencies between the worker's local compute and communication operators, as well as any temporal dependencies specified by Order rules.
The centralized scheduler resolves the ordering between any \chunks and \comms that do not depend on each other (i.e. there is no path between them) and that are assigned the same stream.
For example, these can originate from the nested filters in the \order directive, which indicates the user's intent to overlap sub-DAGs.

We describe the algorithm for scheduling across independent \chunks and \comms.
For simplicity, we use \emph{task} to refer to either a \chunk or \comm.
To decide how to interleave two sub-DAGs, the scheduler begins by creating one queue per stream and initializing the set of ready tasks, which are tasks with no upstream dependencies.
Then, \sys{} performs the following resource scheduling algorithm:

\begin{compactenum}
    \item Pick the ready task $t$ (all upstream tasks scheduled) with the most downstream dependencies.
    \item Add the task to the queue corresponding to $t.stream$.
    \item Mark the task as ready to unblock downstream adjacent tasks.
\end{compactenum}

This simple policy works well for the DualPipe-style overlapping depicted in \Cref{fig:microbatch}b where the forward-backward microbatches are symmetric, meaning that each has the same number and pattern of \comms and \chunks.
It could perform poorly in cases where there are many small tasks in a subgraph that is not on the critical path, or in cases where critical and non-critical path \comms are on the same stream (\Cref{fig:interference}b).
% it could fail to avoid delaying all-to-alls without runtime estimations.  
For example, we could leverage profile-guided optimization to measure actual run times and resource usage.

After this process, each worker will have an ordered list of \chunks and \comms per stream, plus all cross-stream data and temporal dependencies.
Then, once before execution, the centralized scheduler dispatches each device's sub-DAG to the corresponding worker.
The dispatch to a worker also includes the model weights to load for \chunks and \comms.

\subsubsection{Worker execution}

Each worker loads its shard of model weights then executes the compute and communication operators according to the order determined by the centralized scheduler.
The worker is responsible for managing the GPU's local resources that may be shared between tasks: GPU streams, communicators, and memory.
We rely on the GPU to allocate the shared physical resources such as SMs and HBM or network bandwidth between concurrent kernels.
Our approach is complementary to other approaches that provide more precise control over physical resources~\cite{dynaflow,deepep2025}, e.g., adding an SM limit on specific \chunks.

The worker's scheduling loop repeats the following:
\begin{compactenum}
    \item Pick a task \emph{t} from the tasks whose dependencies have been scheduled.
    \item If \emph{t.stream} is different from a dependency's stream \emph{t.args[i].stream}, synchronize by having \emph{t.stream} wait on \emph{t.args[i].event}.
    \item Dispatch \emph{t} to \emph{t.stream}. This launches all kernels in the task's \texttt{exec} function.
    \item Free any of \emph{t.args} that have no more downstream tasks. Store \emph{t.outputs} in local intermediate storage for later consumption. If \emph{t} has an output with a downstream task that runs on a different stream, record an event on \emph{t.stream} and store in \emph{t.outputs[i].event}.
\end{compactenum}

% In this way, the runtime realizes a global execution plan through mechanical local scheduling that is constrained by the global DAG dependencies and resource assignments given by the user.

Note that when using multiple streams, there will often be multiple tasks that are ready for dispatch.
This is because the compiler produces a total ordering of tasks per stream, but only a partial ordering across streams based on data and temporal dependencies. 
When tasks could be scheduled in any order, the worker must make a local decision about dispatch order. 

Although individual kernels execute asynchronously on the GPU, the dispatch order of tasks is still important because dispatch time on the CPU can delay critical GPU kernels, especially when there are many kernels to dispatch in a single task.
For example, when using gradient sharding in ZeRO-2~\cite{zero-deepspeed}, prioritizing gradient reduction could free memory sooner by shortening the lifetime of full gradient state.
On the other hand, prioritizing all-to-all communication over gradient reductions may reduce latency along the critical path.
\sys{} currently prioritizes send communication first, defers receive communication last to reduce point-to-point communication interference, and among remaining communication tasks prioritizes critical-path over reduction operators.
The prioritization is deterministic, to ensure that all ranks in a collective group dispatch communications in the same order.
Future work could include other heuristics and/or online profiling to more precisely determine the best dispatch order.

\vspace{-10pt}

\paragraph{Stream management}
As shown in the above scheduling loop, the worker dispatches each task’s operator(s) on the stream associated with that task’s logical resource. Tasks are dispatched asynchronously, as soon as their upstream tasks have been scheduled, to avoid blocking the GPU on CPU scheduling. When two adjacent tasks execute on different streams, \sys{} inserts synchronization between their streams to preserve correctness using CUDA events and stream-wait. This lets the runtime support arbitrary execution patterns without hard-coding a fixed mapping from operator type to stream.

For example, for overlapping PP communication, send and receive operators should be assigned a different stream from their upstream and downstream compute operators. In this case, the communication stream must synchronize with the producer node’s stream before issuing the send, and the downstream compute node must synchronize with the receive node’s stream before consuming the communicated value.

Although \sys{} must insert synchronization to preserve correctness, it inserts synchronization only when required by inter-task dependencies across different streams, allowing tasks without data dependencies running on independent resources to proceed concurrently.
This also minimizes CPU overhead from scheduling and CUDA event creation.

% When the user specifies that two chunks on the same device should overlap, the actual execution varies based on the user's stream assignment.
% If the user specifies a different chunk per stream, \sys{} simply assigns them to their respective streams to run concurrently with no synchronization in between; this is a simple approach but could lead to unpredictable performance due to resource contention.
% If the chunks share the same set of streams, \sys{} will share the streams between the two chunks by assigning tasks from each.\swang{note about order of assignment, or will that go before the runtime section?}

\vspace{-6pt}

\paragraph{Communication management} Avoiding bubbles from unnecessary synchronization can be challenging due to complex interactions between inter- and intra-device dependencies.
GPU collective communication typically requires a ``communicator'', which represents one rank's context in a given collective group.
Communication operations must be executed in the same order on all ranks' communicators in the same group, or else the device may hang because GPU collective communications are synchronous~\cite{nccl}.
% This is because GPU collective communications are synchronous, meaning that all participants need to enter the same operation before any can continue.

The inter-device communication order can create bubbles when combined with the intra-device dependencies created by shared CUDA streams.
For example, in PP, most workers are continually sending and receiving microbatches to and from another worker.
Thus, if we use one collective group for a PP schedule, then the send and receive operations need to be globally ordered across all workers' communicators.
However, choosing an optimal serial order at compile time is hard because it would require exact predictions of the execution time per microbatch stage as well as communication time.
For example, when choosing between sending and receiving, a worker needs to choose whether to send first, which would unblock the downstream worker's compute, or receive first, which would unblock communication operations queued after the upstream worker's send.

To resolve this problem, \sys{} allocates separate streams for point-to-point communications, one for sending and one for receiving.
To avoid deadlock, \sys{} also allocates one communicator per stream.
Using separate communicators reduces scheduling burden. Instead of requiring that all P2P operations are ordered consistently for every pair of workers, the system only needs to guarantee that P2P operations in each direction are ordered consistently for every pair of workers.
This does not require knowing actual run times and is already naturally satisfied by common PP schedules.
In particular, the requirement on the PP schedule is that downstream workers process data in the same order that it is produced by upstream workers, e.g., if rank 0 executes microbatch 1 before 2, then rank 1 also executes microbatch 1 before 2.
% Currently, \sys{} asks the user to ensure this.
\sys{} currently rejects schedules that do not meet this requirement.

\paragraph{Memory management}
\sys{} manages GPU memory storage for all model state buckets along with all activations sent between \chunks.
Thus, GPU memory is shared by compute \chunks as well as \comms that gather or reduce model state during execution.
For each bucket, \sys{} allocates one flat buffer each for parameters and gradients, concatenating all parameter and gradient tensors in the bucket into their respective flat tensor; copies to/from the flat buffer are elided where possible.

\sys{} also manages memory for ZeRO-style~\cite{zero-deepspeed} gradient/parameter sharding where states must be materialized before computation and sharded after.
In such cases, \sys{} uses persistent buffers to store the sharded states and allocates temporary full buffers for the rematerialized states.
For example, in ZeRO-2 gradient sharding, \sys{} stores persistent buffers for gradient shards and allocates a temporary full gradient buffer at the beginning of each backward task.
% The buffer is released after its last downstream task completes, in this case a reduce-scatter for gradient reduction.
% In ZeRO-3 weight sharding, \sys{} allocates a temporary full weight buffer at the beginning of an all-gather task and releases the buffer at the end of the downstream compute task. 
% Note that the CPU is responsible for freeing temporary buffers.
% Thus, in cases where the last task to consume a buffer executes on a stream other than the stream where it was created, \sys{} uses a background thread to wait on the task to complete and free the temporary buffer's storage.
To precisely control buffer release, \sys{} waits in a background thread for the event corresponding to the last consumer task's completion, then resizes the buffer's storage to zero.
% instead of relying on PyTorch's built-in reference counting for tensors because the autograd engine keeps long-lived references to gradient buffers, making them impossible to free at a precise time.

\sys{} also manages intermediate tensors that will later be consumed by downstream tasks. These include in-flight forward activations that must be retained for later backpropagation, activations waiting to be sent to a downstream worker, and intermediate autograd tensors that will later be used to compute model gradients. Once the last consumer task is scheduled, \sys{} frees the corresponding tensor.
\section{Implementation}

\sys{} is implemented as a TorchDynamo~\cite{pytorch2} backend, allowing it to hook into arbitrary PyTorch code.
% TorchDynamo is a PyTorch JIT compiler which uses symbolic tracing to capture PyTorch operator graphs, called \texttt{fx.Graph}s.
Although TorchDynamo supports non-static behavior by combining compiled- and eager-model execution, \sys{} currently requires models to be fully traceable.
This allows \sys{} to partition an \texttt{fx.Graph} into sub-graphs at compile time. 
% Each sub-\texttt{fx.Graph} serves as the executable representation for compute \chunks.
% The PyTorch autograd engine implicitly adds the executable graph for backwards \chunks. 
Annotations are implemented as Python context managers that attach metadata to regions of the model during execution. During graph capture, these annotations are recorded and used to segment the \texttt{fx.Graph} into \chunks.

\sys{}'s runtime is built on top of Ray.
The \sys{} compiler and centralized scheduler~(\Cref{sec:design:runtime}) run together in a ``driver'' process. Each worker is implemented as a Ray actor.
\section{Evaluation}

We evaluate \sys{} against the general-purpose training frameworks Megatron-LM 0.18.0~\cite{megatron-nvidia}, DeepSpeed 0.18.9~\cite{deepspeed-microsoft} and TorchTitan 0.2.2~\cite{liang2025torchtitan}.
We address the following questions.
\begin{enumerate}
    \item Does \sys{} perform as well as existing systems on commonly supported strategies?
    \item What benefits in strategy flexibility and performance does \sys{} provide? Here, we evaluate performance and memory efficiency of various combinations of PP and ZeRO~\cite{zero-deepspeed}, along with PP and EP, including the DualPipeV schedule~\cite{qi2025dual}.
    \item How well does \sys{} scale?
\end{enumerate}

We run our evaluation on 4 AWS EC2 NVIDIA 8xA100 nodes with NVLink intra-node interconnect and Elastic Fabric Adapter (EFA) enabled for inter-node networking. 

\subsection{Common PP strategies}
We evaluate \sys{}'s support for common PP schedules against Megatron and TorchTitan, which both provide out-of-the-box support for the 1F1B schedule~\cite{1f1b-msr} and its interleaved variant~\cite{i1f1b-nvidia}.
Interleaved 1F1B can improve training throughput for large models compared to 1F1B due to better microbatch overlapping from finer-grained interleaved stages.

TorchTitan allows users to provide a PP schedule, similar to \sys{}, and provides generic PP schedule-builders for 1F1B and interleaved 1F1B.
We adapt the 1F1B and interleaved 1F1B schedule builders from TorchTitan to the \sys{} API in 29 LoC and 38 LoC, respectively.

We distribute the Qwen3 1B model with PP-8 x DP/EP-4 across 32 A100 GPUs and the Qwen3 9B model with PP-4 x DP/EP-4 across 16 A100 GPUs.
We make DP the outer dimension to simulate larger-scale settings where cross-node DP is required. 
Megatron requires placing PP as the outermost dimension, so we route intra-node traffic over PCIe to simulate lower inter-node bandwidth for Megatron.

% For \sys{} and TorchTitan, we report 1F1B and interleaved 1F1B results with and without composition with TorchInductor, TorchDynamo's default backend for \texttt{fx.Graph} optimization. 
% TorchInductor applies optimizations such as kernel fusion to the \texttt{fx.Graph} and produces optimized Triton~\cite{xxx} GPU kernels.
% This approach is complimentary to \sys{}; we show that within-\chunk optimizations like Inductor can be composed easily. 

TorchTitan's performance suffers due to a larger memory footprint compared to the other systems, leading to CPU-side delays inserted by the PyTorch CUDA allocator.
The memory footprint is caused by memory buffers needed by TorchTitan's DP implementation.
Additionally, TorchTitan's interleaved schedule does 14\% worse than its own 1F1B schedule.
This is due to sends and recvs executing on the same stream, creating bubbles across ranks in the interleaved schedule.

Meanwhile, \sys{}-interleaved-1F1B achieves 5\% higher throughput than \sys{}-1F1B, as expected.
This is due to a lower memory footprint and use of dual P2P streams and communicators, one for sending and one for receiving.

\sys{}'s performance is comparable with Megatron's.
% particularly when \sys{} enables TorchInductor to compile operator subgraphs~\cite{pytorch2}.
We note that Megatron often uses hand-tuned fused kernels targeted to transformer operators, enabling faster single-device execution time: on a single device, one microbatch forward pass of a single PP stage takes about 30 ms in Megatron vs 40 ms in \sys{} for Qwen3 1B.
In general, \sys{} is orthogonal to approaches like kernel fusion which apply to the low-level operator graph contained within a \sys{} \chunk.\texttt{exec} function.
Thus, we believe that the two approaches could be combined to further improve performance gains.

\begin{figure}
    \centering
    \includegraphics[width=\columnwidth]{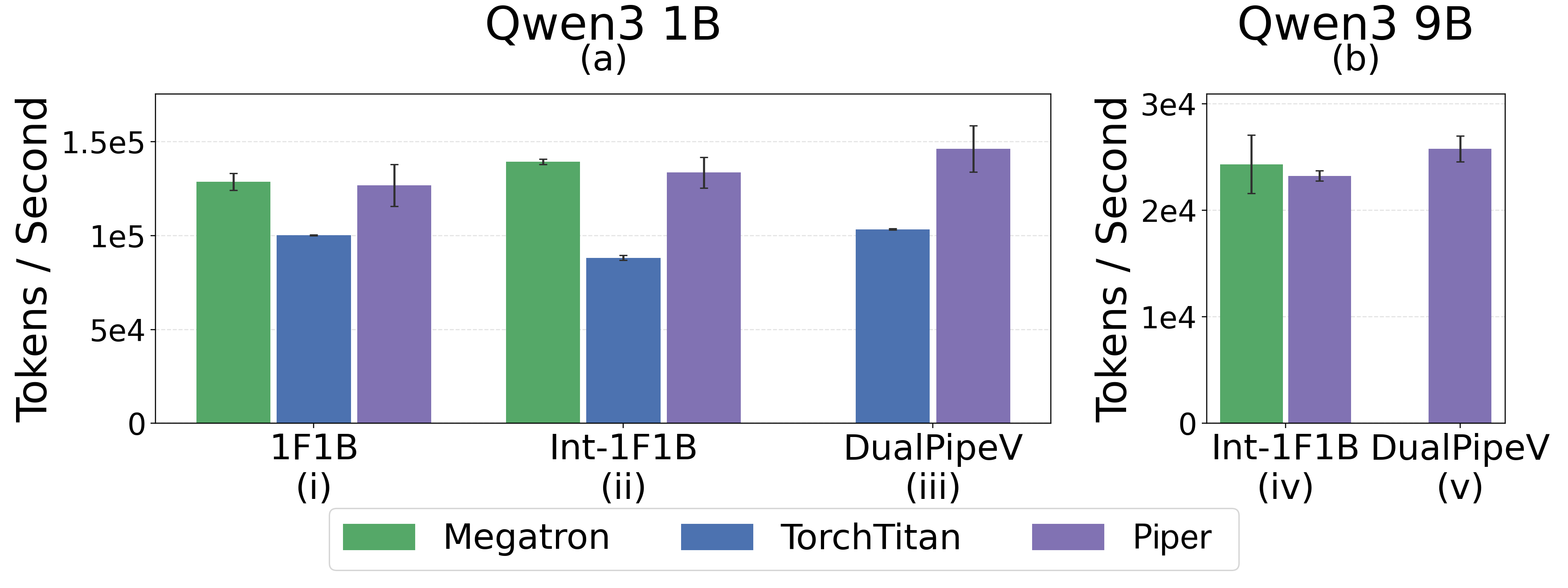}
    \caption{PP x EP throughput for 1F1B, Interleaved 1F1B and DualPipeV schedules.}
    \label{fig:schedules}
\end{figure}

\vspace{-10pt}

\begin{table}[t]
\centering
\begin{tabular}{l c c c}
\hline
Framework & ZeRO-1 & ZeRO-2 & ZeRO-3 \\
\hline
DeepSpeed    & \cmark & \xmark & \xmark \\
Megatron-LM  & \cmark & \xmark & \xmark \\
TorchTitan   & \cmark & \astmark & \astmark \\
\sys        & \cmark & \cmark & \cmark \\
\hline
\end{tabular}
\caption{PP x ZeRO support in the evaluated systems.
TorchTitan claims to support ZeRO-2 and ZeRO-3 composed with PP~\cite{liang2025torchtitan}, but we found that in practice gradient and weight states do not get resharded between all microbatches, so the memory savings is significantly less than expected.
\sys{} supports all combinations of PP with ZeRO memory optimizations.
}
\label{tab:zero-support}
\end{table}

\vspace{-10pt}

\begin{table}[t]
\centering
\begin{tabular}{r c c c c}
\hline
Framework & \sys{} & TT & DeepSpeed & Megatron \\
Tokens/s  & 8641  & 8637 & 9352      & 9942     \\
$\pm$     & 701   & 977  & 52        & 1106     \\
\hline
\end{tabular}
\caption{DP ZeRO-1 throughput on all systems.}
\label{tab:zero1-throughput}
\end{table}

\subsection{PP x ZeRO}
\label{sec:eval:zero}

\Cref{tab:zero-support} shows the current support for PP x ZeRO variations in Megatron-LM, DeepSpeed and TorchTitan.
All systems support PP x ZeRO-1, which is an optimization of DP where redundant optimizer state is deduplicated across DP replicas.

To evaluate basic PP x ZeRO-1 support in all systems, we distribute the Qwen3 1B model with DP-2 across 2 A100 GPUs and evaluate training throughput with ZeRO-1 optimizer state sharding. 
~\Cref{tab:zero1-throughput} shows that all systems perform similarly, as expected.
Differences in execution time are likely due to differences in low-level kernel implementations.

While all systems support ZeRO variants 1-3, the support for ZeRO 2 (parameter sharding) and 3 (weights sharding) when combined with PP is incomplete.
Megatron and DeepSpeed do not support either when combined with PP.
TorchTitan supports the variants but the rematerialized buffers are not resharded between all PP microbatches.
This reduces communication overhead but also defeats the purpose of ZeRO sharding, which is to save per-device memory from redundant states.
We believe that this uneven support for ZeRO-2 and ZeRO-3 is due to the fact that gradient and weights sharding requires careful management of pre- and post-layer hooks, whereas ZeRO-1 only operates on the optimizer states.
% TorchTitan supports all three with some limitations for model sharding.
Meanwhile, \sys{} correctly supports all combinations of PP x ZeRO; it simply adds the proper allgather, reshard, and reduce-scatter operations to the matched \chunks in the DAG IR, irrespective of the PP strategy.

We validate the implications of this compatibility matrix by evaluating peak memory of composed PP x ZeRO strategies.
We distribute Qwen3 9B with 8-way PP and 4-way DP across 32 A100 GPUs.
We expect peak training memory to decrease from ZeRO-2 to ZeRO-3, as the latter introduces parameter sharding in addition to gradient state sharding.
% \megan{by X\%? its 25\% just considering model state (grads fp32, params fp16)}

\begin{figure}
    \centering
    \includegraphics[width=\columnwidth]{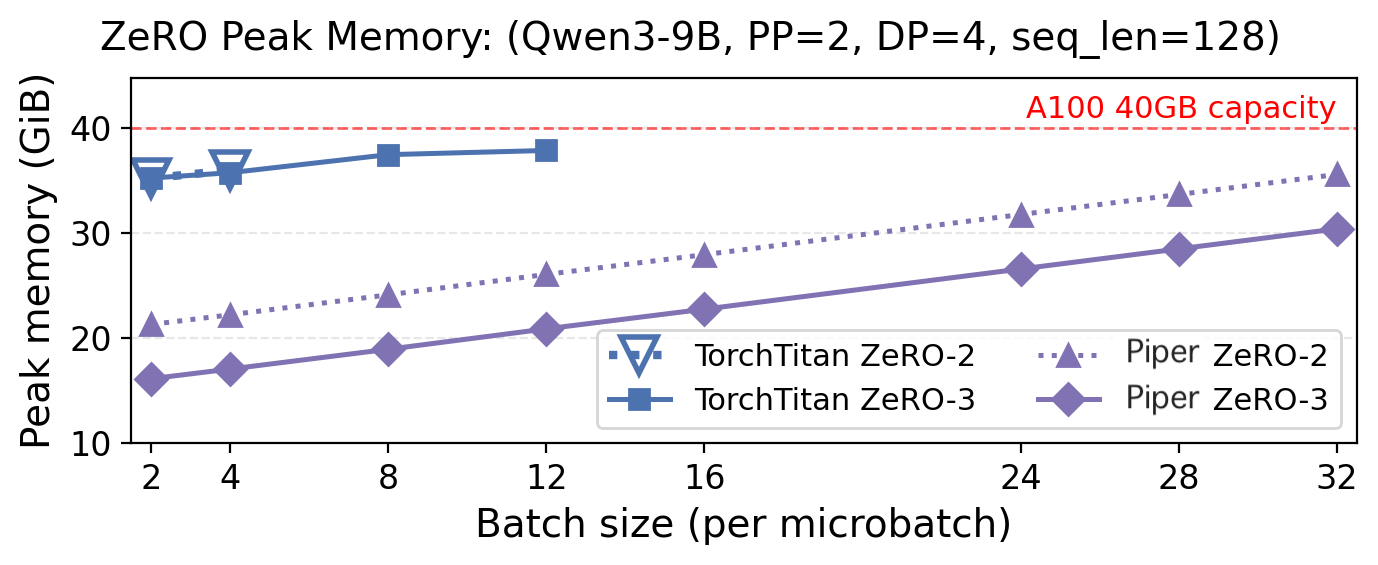}
    \caption{Peak Memory consumption of ZeRO-2 and ZeRO-3 sharding strategies for \sys and TorchTitan.}
    \label{fig:zero-memory-sweep}
\end{figure}

\Cref{fig:zero-memory-sweep} shows the results.
Megatron, DeepSpeed, and other PP x ZeRO-1 variants OOM in all cases because they cannot fit even the smallest batch size.
TorchTitan can fit the smaller batch sizes but has higher memory consumption than \sys{} because it does not properly re-shard weights and gradients, despite setting the provided flag for weights re-sharding after forwards.
As a result, TorchTitan maintains a full copy of parameters and gradients across multiple microbatches
% TorchTitan offers a flag to force re-sharding after forwards microbatches, but we observed that the full parameters are still kept alive through consecutive backward stages.
and OOMs at batch size 8 for ZeRO-2 and batch size 16 for ZeRO-3.
TorchTitan's PP x ZeRO-3 implementation also does not save significant memory compared to ZeRO-2.

For \sys{}, we achieve correct resharding by reducing gradients after every backward pass instead of accumulating gradients between PP microbatches.
% We set the bucket size to 25MB to ensure that few enough weights/gradients are simultaneously in memory.
This adds communication time compared to TorchTitan but achieves the full ZeRO memory savings.
As a result, \sys{} is able to execute with much higher batch sizes than TorchTitan without OOM: 8x higher for ZeRO-2 (up to batch size 32) and 3.3x higher for ZeRO-3 (up to batch size 40).
% \sys avoids OOMs until batch size XXX for ZeRO-2 and batch size X for ZeRO-3.

% \begin{figure*}
%     \centering
%     \includegraphics[width=\textwidth]{figures/zero1_dp_memory_compare.png}
%     \caption{Memory consumption of ZeRO-1 sharding strategy for \sys and TorchTitan.}
%     \label{fig:zero1-dp-memory-sweep}
% \end{figure*}

% \begin{figure}
%     \centering
%     \includegraphics[width=\columnwidth]{figures/zero1_dp_step_time_compare.png}
%     \caption{Iteration times in ZeRO-1 sharding strategy for \sys and TorchTitan, for batch size of 32.\megan{can include these numbers in the text to save space}\swang{change this to throughput}}
%     \label{fig:zero1-time-dp-sweep}
% \end{figure}

\vspace{-10pt}
  
\subsection{PP x EP and DualPipe}
We compare system support for DualPipe-like schedules across Megatron, TorchTitan, and \sys{}.
Megatron provides custom schedules for 1F1B and interleaved 1F1B but not DualPipe, likely due to the need to overlap between microbatches.
TorchTitan allows users to provide a PP schedule, similar to \sys{}, and provides a DualPipeV schedule builder~\cite{qi2025dual}.
DualPipeV (and DualPipe) also uses interleaving, so we use each system's interleaved-1F1B implementation as its respective baseline; the expected performance gain in each case is from overlapping of some EP communication.
TorchTitan's DualPipeV implementation uses separate threads to dispatch the overlapped forward and backward microbatches to two streams, one for compute and one for EP communication.

We adapt the DualPipeV schedule-builder from TorchTitan to the \sys{} API in 63 LoC.
We use the \texttt{Order} directive to specify microbatch overlapping.
Similar to TorchTitan, the runtime also uses a background stream for EP communication, but the tasks from the overlapped microbatches are scheduled jointly by a single thread.

We use the same experiment setup as in Figure \ref{fig:schedules}a.
In the 1B setting, \sys{}-DualPipeV improves 13\% over \sys{}-1F1B schedule, while TorchTitan-DualPipeV improves only 3\% over TorchTitan-1F1B.
We believe that TorchTitan's lack of improvement is due to a lack of synchronization between the forwards and backwards dispatch threads, which leads to unintended serialization of the GPU's compute and communication streams.

For the 9B model, we were not able to run TorchTitan due to OOM.
\sys{}-DualPipeV improves 10\% over its interleaved schedule and 6\% over Megatron's interleaved schedule~(\Cref{fig:schedules}b.
In this case, Megatron-Interleaved-1F1B's improvement over \sys{}-Interleaved-1F1B comes from fused kernels.
Note that Megatron does not fuse the EP all-to-all; thus we expect that Megatron's fused kernels could be directly integrated into \sys{} for additional gains.

\subsection{Scalability}
We evaluate \sys{}'s PP and DP scalability by distributing Qwen3 1B with 2-, 4- and 8-way PP and 2- and 4-way DP
We scale the global batch size linearly with the PP and DP degrees and plot the curve representing linear scaling. 
We show that \sys{} scales reasonably.

\begin{figure}
    \centering
    \includegraphics[width=\columnwidth]{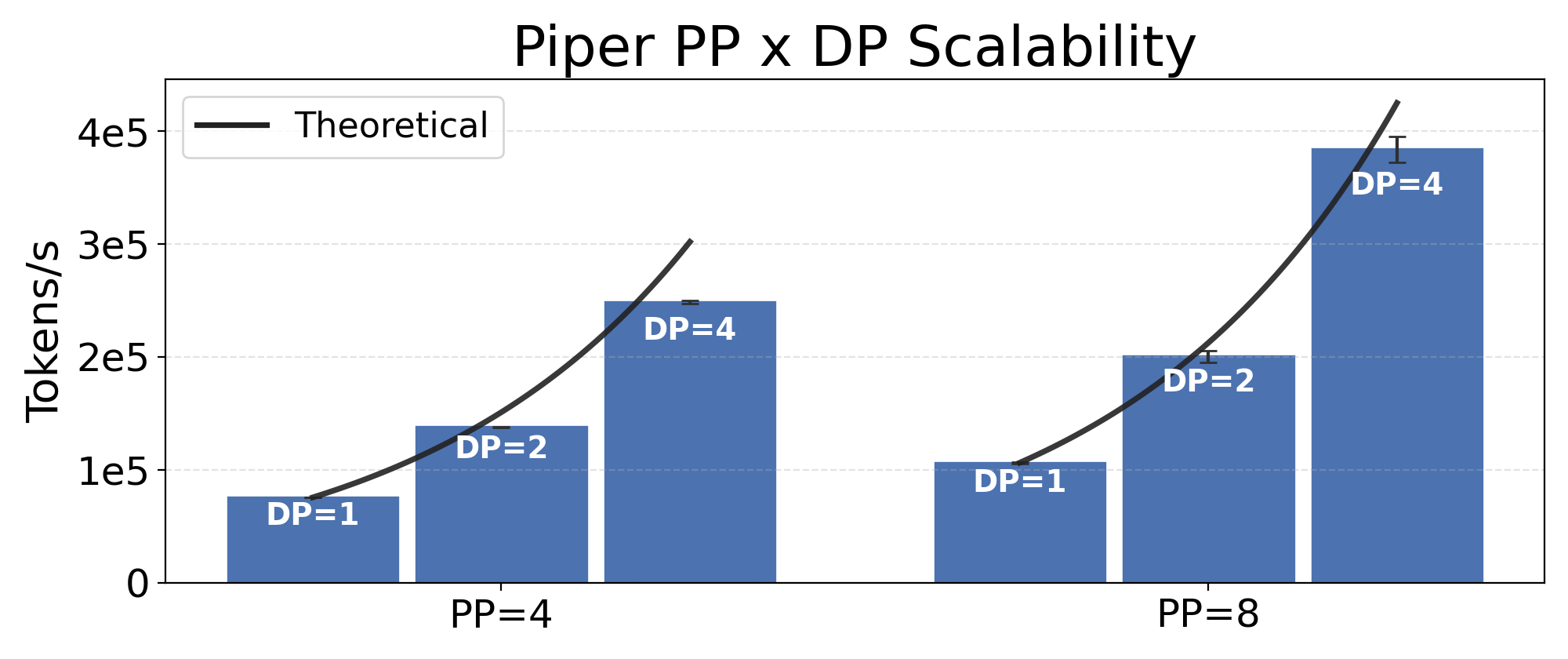}
    \caption{\sys{} PP x DP Scalability.}
    \label{fig:placeholder}
\end{figure}
\section{Related Work}

\paragraph{General-purpose training frameworks}
% PyTorch's eager execution model makes it challenging to build a fully flexible distributed training runtime.
Most general-purpose training frameworks~\cite{megatron-nvidia,deepspeed-microsoft,liang2025torchtitan} for PyTorch implement a fixed set of parallelism strategies.
DP/ZeRO strategies are model-agnostic and implemented with pre- and post-layer hooks.
Megatron-LM and DeepSpeed implement custom-built layers for TP/EP/CP and schedules for PP.
TorchTitan supports more general strategies by using DTensor~\cite{dtensor} for sharding, an API inspired by GSPMD~\cite{gspmd-google}, and PiPPy~\cite{pippy2022} for PP schedules.
However, in all systems, each parallelism dimension dispatches its compute and communication operations eagerly with little to no synchronization with other operations, making it difficult to jointly schedule shared GPU resources such as communication bandwidth or memory.

Compiler-based systems such as JAX/XLA~\cite{bradbury2021jax} provide high-level tensor sharding and replication annotations based on GShard~\cite{lepikhin2020gshard} and GSPMD~\cite{gspmd-google}.
\sys{}'s API is inspired by these works.
The annotations allow the compiler to insert the necessary communications and lower to a per-device program, although it is limited to homogeneous SPMD strategies.
Later work extended the API and runtime to support arbitrary PP schedules~\cite{jaxpp-nvidia} and heterogeneous strategies or clusters~\cite{pathways-google}.
However, JAX/XLA does not support user scheduling of the low-level execution strategy; per-device resources such as GPU streams are opaque to the user and changes to the low-level execution strategy would require modifications deep within the XLA compiler.

\paragraph{DSLs}
Besides GSPMD, many other works have combined user-facing tensor annotations with compiler support to enable flexible distributed training strategies.
These include CoCoNeT~\cite{jangda2022breaking} for automatic computation-communication fusion, AutoSP~\cite{guptaautosp} for automatic sequence parallelism, and DynaFlow for controlling intra-device parallelism~\cite{dynaflow}.
Integration of these works is an interesting line of future work.

Slapo~\cite{chen2024slapo} shares a similar motivation of user scheduling for distributed training.
However, it does not consider intra-device parallelism and lacks a flexible runtime, offloading execution to general-purpose frameworks~\cite{megatron-nvidia,deepspeed-microsoft}.
TVM~\cite{tvm}, inspired by Halide~\cite{halide}, supports user scheduling and autotuning of tensor programs; we are inspired by these works and extend them to support distributed tensor programs.

\paragraph{Auto-parallelism}
There have been numerous systems that aim to automatically find an optimal strategy~\cite{lai2023merak,liu2024aceso,miao2022galvatron,sun2024adapipe,tarnawski2021piper,wang2019supporting,yuan2024accelerating,jia2019beyond,unger2022unity,alpa-ucb,lin2024nnscaler,zhu2025mist}.
To make the search problem tractable, many of these systems focus on a subset of strategies.
They may not support execution of strategies that fall outside the search space and may thus miss opportunities such as fully overlapping communication and computation, as noted by \cite{zhu2025mist}.
ByteScheduler~\cite{peng2019generic} and Centauri~\cite{chen2024centauri} do consider communication scheduling and Centauri additionally provides communication partitioning primitives, similar to \sys{}.
% Mist searches over multiple parallelism strategies jointly with memory optimizations such as ZeRO.
However, in general, these systems are still limited to the dimensions chosen by their developers and lack a unified IR for jointly scheduling the operations inserted by composed high-level strategies, such as in DualPipe.
We hope \sys{} can serve as a common runtime for these auto-parallelism systems.

nnScaler~\cite{lin2024nnscaler} is the closest in approach. It allows the user to specify generic constraints that are similar to \sys{} directives.
However, it does not support intra-device parallelism.

Many systems use profile-guided optimization to identify a good strategy.
Notable examples include DeepCompile~\cite{tanaka2025deepcompile} which searches over ZeRO and other memory optimizations, and Tessera~\cite{tessera-osdi26} which co-optimizes the PP partitioning and microbatch overlapping schedule and dynamically schedules backwards passes.
In future work, we also plan to integrate these works' dynamic approaches.
\section{Conclusion}

We present \sys{}, a distributed training system for PyTorch that decouples the distributed execution strategy specification from the model and runtime using a unified IR: the global training DAG. 
By making the placement, granularity, and ordering of computation and communication operations explicit, \sys{} allows users to express a wide range of distributed training strategies combining PP, DP, and EP with arbitrary ZeRO memory optimizations. 
\sys{} provides a useful interface for encoding both expert-designed schedules and future automated search over a rich space of composed strategies.

\bibliographystyle{ACM-Reference-Format}
\bibliography{src/references}


\begin{thebibliography}{57}

%%% the Web address.

\ifx \showCODEN    \undefined \def \showCODEN     #1{\unskip}     \fi
\ifx \showISBNx    \undefined \def \showISBNx     #1{\unskip}     \fi
\ifx \showISBNxiii \undefined \def \showISBNxiii  #1{\unskip}     \fi
\ifx \showISSN     \undefined \def \showISSN      #1{\unskip}     \fi
\ifx \showLCCN     \undefined \def \showLCCN      #1{\unskip}     \fi
\ifx \shownote     \undefined \def \shownote      #1{#1}          \fi
\ifx \showarticletitle \undefined \def \showarticletitle #1{#1}   \fi
\ifx \showURL      \undefined \def \showURL       {\relax}        \fi
% invisible to TeX
\providecommand\bibfield[2]{#2}
\providecommand\bibinfo[2]{#2}
\providecommand\natexlab[1]{#1}
\providecommand\showeprint[2][]{arXiv:#2}

\bibitem[dee(2025)]%
        {deepspeed-microsoft}
 \bibinfo{year}{2025}\natexlab{}.
\newblock \bibinfo{title}{DeepSpeed: Extreme-scale model training for everyone}.
\newblock \bibinfo{howpublished}{\url{https://www.microsoft.com/en-us/research/blog/deepspeed-extreme-scale-model-training-for-everyone}}.
\newblock


\bibitem[dte(2025)]%
        {dtensor}
 \bibinfo{year}{2025}\natexlab{}.
\newblock \bibinfo{title}{torch.distributed.tensor}.
\newblock \bibinfo{howpublished}{\url{https://docs.pytorch.org/docs/stable/distributed.tensor.html}}.
\newblock
\newblock
\shownote{Package}.


\bibitem[Abadi et~al\mbox{.}(2016)]%
        {tensorflow}
\bibfield{author}{\bibinfo{person}{Mart\'{\i}n Abadi}, \bibinfo{person}{Paul Barham}, \bibinfo{person}{Jianmin Chen}, \bibinfo{person}{Zhifeng Chen}, \bibinfo{person}{Andy Davis}, \bibinfo{person}{Jeffrey Dean}, \bibinfo{person}{Matthieu Devin}, \bibinfo{person}{Sanjay Ghemawat}, \bibinfo{person}{Geoffrey Irving}, \bibinfo{person}{Michael Isard}, \bibinfo{person}{Manjunath Kudlur}, \bibinfo{person}{Josh Levenberg}, \bibinfo{person}{Rajat Monga}, \bibinfo{person}{Sherry Moore}, \bibinfo{person}{Derek~G. Murray}, \bibinfo{person}{Benoit Steiner}, \bibinfo{person}{Paul Tucker}, \bibinfo{person}{Vijay Vasudevan}, \bibinfo{person}{Pete Warden}, \bibinfo{person}{Martin Wicke}, \bibinfo{person}{Yuan Yu}, {and} \bibinfo{person}{Xiaoqiang Zheng}.} \bibinfo{year}{2016}\natexlab{}.
\newblock \showarticletitle{TensorFlow: a system for large-scale machine learning}. In \bibinfo{booktitle}{\emph{Proceedings of the 12th USENIX Conference on Operating Systems Design and Implementation}} (Savannah, GA, USA) \emph{(\bibinfo{series}{OSDI'16})}. \bibinfo{publisher}{USENIX Association}, \bibinfo{address}{USA}, \bibinfo{pages}{265–283}.
\newblock
\showISBNx{9781931971331}


\bibitem[Ansel et~al\mbox{.}(2024)]%
        {pytorch2}
\bibfield{author}{\bibinfo{person}{Jason Ansel}, \bibinfo{person}{Edward Yang}, \bibinfo{person}{Horace He}, \bibinfo{person}{Natalia Gimelshein}, \bibinfo{person}{Animesh Jain}, \bibinfo{person}{Michael Voznesensky}, \bibinfo{person}{Bin Bao}, \bibinfo{person}{Peter Bell}, \bibinfo{person}{David Berard}, \bibinfo{person}{Evgeni Burovski}, \bibinfo{person}{Geeta Chauhan}, \bibinfo{person}{Anjali Chourdia}, \bibinfo{person}{Will Constable}, \bibinfo{person}{Alban Desmaison}, \bibinfo{person}{Zachary DeVito}, \bibinfo{person}{Elias Ellison}, \bibinfo{person}{Will Feng}, \bibinfo{person}{Jiong Gong}, \bibinfo{person}{Michael Gschwind}, \bibinfo{person}{Brian Hirsh}, \bibinfo{person}{Sherlock Huang}, \bibinfo{person}{Kshiteej Kalambarkar}, \bibinfo{person}{Laurent Kirsch}, \bibinfo{person}{Michael Lazos}, \bibinfo{person}{Mario Lezcano}, \bibinfo{person}{Yanbo Liang}, \bibinfo{person}{Jason Liang}, \bibinfo{person}{Yinghai Lu}, \bibinfo{person}{C.~K. Luk}, \bibinfo{person}{Bert Maher}, \bibinfo{person}{Yunjie
  Pan}, \bibinfo{person}{Christian Puhrsch}, \bibinfo{person}{Matthias Reso}, \bibinfo{person}{Mark Saroufim}, \bibinfo{person}{Marcos~Yukio Siraichi}, \bibinfo{person}{Helen Suk}, \bibinfo{person}{Shunting Zhang}, \bibinfo{person}{Michael Suo}, \bibinfo{person}{Phil Tillet}, \bibinfo{person}{Xu Zhao}, \bibinfo{person}{Eikan Wang}, \bibinfo{person}{Keren Zhou}, \bibinfo{person}{Richard Zou}, \bibinfo{person}{Xiaodong Wang}, \bibinfo{person}{Ajit Mathews}, \bibinfo{person}{William Wen}, \bibinfo{person}{Gregory Chanan}, \bibinfo{person}{Peng Wu}, {and} \bibinfo{person}{Soumith Chintala}.} \bibinfo{year}{2024}\natexlab{}.
\newblock \showarticletitle{PyTorch 2: Faster Machine Learning Through Dynamic Python Bytecode Transformation and Graph Compilation}. In \bibinfo{booktitle}{\emph{Proceedings of the 29th ACM International Conference on Architectural Support for Programming Languages and Operating Systems, Volume 2}} (La Jolla, CA, USA) \emph{(\bibinfo{series}{ASPLOS '24})}. \bibinfo{publisher}{Association for Computing Machinery}, \bibinfo{address}{New York, NY, USA}, \bibinfo{pages}{929–947}.
\newblock
\showISBNx{9798400703850}
\href{https://doi.org/10.1145/3620665.3640366}{doi:\nolinkurl{10.1145/3620665.3640366}}


\bibitem[Barham et~al\mbox{.}(2022)]%
        {pathways-google}
\bibfield{author}{\bibinfo{person}{Paul Barham}, \bibinfo{person}{Aakanksha Chowdhery}, \bibinfo{person}{Jeff Dean}, \bibinfo{person}{Sanjay Ghemawat}, \bibinfo{person}{Steven Hand}, \bibinfo{person}{Dan Hurt}, \bibinfo{person}{Michael Isard}, \bibinfo{person}{Hyeontaek Lim}, \bibinfo{person}{Ruoming Pang}, \bibinfo{person}{Sudip Roy}, \bibinfo{person}{Brennan Saeta}, \bibinfo{person}{Parker Schuh}, \bibinfo{person}{Ryan Sepassi}, \bibinfo{person}{Laurent~El Shafey}, \bibinfo{person}{Chandramohan~A. Thekkath}, {and} \bibinfo{person}{Yonghui Wu}.} \bibinfo{year}{2022}\natexlab{}.
\newblock \bibinfo{title}{Pathways: Asynchronous Distributed Dataflow for ML}.
\newblock
\showeprint[arxiv]{2203.12533}~[cs.DC]
\urldef\tempurl%
\url{https://arxiv.org/abs/2203.12533}
\showURL{%
\tempurl}


\bibitem[Bradbury et~al\mbox{.}(2021)]%
        {bradbury2021jax}
\bibfield{author}{\bibinfo{person}{James Bradbury}, \bibinfo{person}{Roy Frostig}, \bibinfo{person}{Peter Hawkins}, \bibinfo{person}{Matthew~James Johnson}, \bibinfo{person}{Chris Leary}, \bibinfo{person}{Dougal Maclaurin}, \bibinfo{person}{George Necula}, \bibinfo{person}{Adam Paszke}, \bibinfo{person}{Jake VanderPlas}, \bibinfo{person}{Skye Wanderman-Milne}, {et~al\mbox{.}}} \bibinfo{year}{2021}\natexlab{}.
\newblock \showarticletitle{Jax: Autograd and xla}.
\newblock \bibinfo{journal}{\emph{Astrophysics Source Code Library}} (\bibinfo{year}{2021}), \bibinfo{pages}{ascl--2111}.
\newblock


\bibitem[Cao et~al\mbox{.}(2026)]%
        {qwen3-coder-next}
\bibfield{author}{\bibinfo{person}{Ruisheng Cao}, \bibinfo{person}{Mouxiang Chen}, \bibinfo{person}{Jiawei Chen}, \bibinfo{person}{Zeyu Cui}, \bibinfo{person}{Yunlong Feng}, \bibinfo{person}{Binyuan Hui}, \bibinfo{person}{Yuheng Jing}, \bibinfo{person}{Kaixin Li}, \bibinfo{person}{Mingze Li}, \bibinfo{person}{Junyang Lin}, \bibinfo{person}{Zeyao Ma}, \bibinfo{person}{Kashun Shum}, \bibinfo{person}{Xuwu Wang}, \bibinfo{person}{Jinxi Wei}, \bibinfo{person}{Jiaxi Yang}, \bibinfo{person}{Jiajun Zhang}, \bibinfo{person}{Lei Zhang}, \bibinfo{person}{Zongmeng Zhang}, \bibinfo{person}{Wenting Zhao}, {and} \bibinfo{person}{Fan Zhou}.} \bibinfo{year}{2026}\natexlab{}.
\newblock \bibinfo{title}{Qwen3-Coder-Next Technical Report}.
\newblock
\showeprint[arxiv]{2603.00729}~[cs.CL]
\urldef\tempurl%
\url{https://arxiv.org/abs/2603.00729}
\showURL{%
\tempurl}


\bibitem[Chang et~al\mbox{.}(2024)]%
        {chang2024flux}
\bibfield{author}{\bibinfo{person}{Li-Wen Chang}, \bibinfo{person}{Wenlei Bao}, \bibinfo{person}{Qi Hou}, \bibinfo{person}{Chengquan Jiang}, \bibinfo{person}{Ningxin Zheng}, \bibinfo{person}{Yinmin Zhong}, \bibinfo{person}{Xuanrun Zhang}, \bibinfo{person}{Zuquan Song}, \bibinfo{person}{Chengji Yao}, \bibinfo{person}{Ziheng Jiang}, {et~al\mbox{.}}} \bibinfo{year}{2024}\natexlab{}.
\newblock \showarticletitle{Flux: Fast software-based communication overlap on gpus through kernel fusion}.
\newblock \bibinfo{journal}{\emph{arXiv preprint arXiv:2406.06858}} (\bibinfo{year}{2024}).
\newblock


\bibitem[Chen et~al\mbox{.}(2024a)]%
        {chen2024centauri}
\bibfield{author}{\bibinfo{person}{Chang Chen}, \bibinfo{person}{Xiuhong Li}, \bibinfo{person}{Qianchao Zhu}, \bibinfo{person}{Jiangfei Duan}, \bibinfo{person}{Peng Sun}, \bibinfo{person}{Xingcheng Zhang}, {and} \bibinfo{person}{Chao Yang}.} \bibinfo{year}{2024}\natexlab{a}.
\newblock \showarticletitle{Centauri: Enabling efficient scheduling for communication-computation overlap in large model training via communication partitioning}. In \bibinfo{booktitle}{\emph{Proceedings of the 29th ACM International Conference on Architectural Support for Programming Languages and Operating Systems, Volume 3}}. \bibinfo{pages}{178--191}.
\newblock


\bibitem[Chen et~al\mbox{.}(2024b)]%
        {chen2024slapo}
\bibfield{author}{\bibinfo{person}{Hongzheng Chen}, \bibinfo{person}{Cody~Hao Yu}, \bibinfo{person}{Shuai Zheng}, \bibinfo{person}{Zhen Zhang}, \bibinfo{person}{Zhiru Zhang}, {and} \bibinfo{person}{Yida Wang}.} \bibinfo{year}{2024}\natexlab{b}.
\newblock \showarticletitle{Slapo: A schedule language for progressive optimization of large deep learning model training}. In \bibinfo{booktitle}{\emph{Proceedings of the 29th ACM International Conference on Architectural Support for Programming Languages and Operating Systems, Volume 2}}. \bibinfo{pages}{1095--1111}.
\newblock


\bibitem[Chen et~al\mbox{.}(2018)]%
        {tvm}
\bibfield{author}{\bibinfo{person}{Tianqi Chen}, \bibinfo{person}{Thierry Moreau}, \bibinfo{person}{Ziheng Jiang}, \bibinfo{person}{Haichen Shen}, \bibinfo{person}{Eddie~Q. Yan}, \bibinfo{person}{Leyuan Wang}, \bibinfo{person}{Yuwei Hu}, \bibinfo{person}{Luis Ceze}, \bibinfo{person}{Carlos Guestrin}, {and} \bibinfo{person}{Arvind Krishnamurthy}.} \bibinfo{year}{2018}\natexlab{}.
\newblock \showarticletitle{{TVM:} End-to-End Optimization Stack for Deep Learning}.
\newblock \bibinfo{journal}{\emph{CoRR}}  \bibinfo{volume}{abs/1802.04799} (\bibinfo{year}{2018}).
\newblock
\showeprint[arXiv]{1802.04799}
\urldef\tempurl%
\url{http://arxiv.org/abs/1802.04799}
\showURL{%
\tempurl}


\bibitem[DeepSeek-AI et~al\mbox{.}(2025)]%
        {deepseekv3-deepseekai}
\bibfield{author}{\bibinfo{person}{DeepSeek-AI}, \bibinfo{person}{Aixin Liu}, \bibinfo{person}{Bei Feng}, \bibinfo{person}{Bing Xue}, \bibinfo{person}{Bingxuan Wang}, \bibinfo{person}{Bochao Wu}, \bibinfo{person}{Chengda Lu}, \bibinfo{person}{Chenggang Zhao}, \bibinfo{person}{Chengqi Deng}, \bibinfo{person}{Chenyu Zhang}, \bibinfo{person}{Chong Ruan}, \bibinfo{person}{Damai Dai}, \bibinfo{person}{Daya Guo}, \bibinfo{person}{Dejian Yang}, \bibinfo{person}{Deli Chen}, \bibinfo{person}{Dongjie Ji}, \bibinfo{person}{Erhang Li}, \bibinfo{person}{Fangyun Lin}, \bibinfo{person}{Fucong Dai}, \bibinfo{person}{Fuli Luo}, \bibinfo{person}{Guangbo Hao}, \bibinfo{person}{Guanting Chen}, \bibinfo{person}{Guowei Li}, \bibinfo{person}{H. Zhang}, \bibinfo{person}{Han Bao}, \bibinfo{person}{Hanwei Xu}, \bibinfo{person}{Haocheng Wang}, \bibinfo{person}{Haowei Zhang}, \bibinfo{person}{Honghui Ding}, \bibinfo{person}{Huajian Xin}, \bibinfo{person}{Huazuo Gao}, \bibinfo{person}{Hui Li}, \bibinfo{person}{Hui Qu},
  \bibinfo{person}{J.~L. Cai}, \bibinfo{person}{Jian Liang}, \bibinfo{person}{Jianzhong Guo}, \bibinfo{person}{Jiaqi Ni}, \bibinfo{person}{Jiashi Li}, \bibinfo{person}{Jiawei Wang}, \bibinfo{person}{Jin Chen}, \bibinfo{person}{Jingchang Chen}, \bibinfo{person}{Jingyang Yuan}, \bibinfo{person}{Junjie Qiu}, \bibinfo{person}{Junlong Li}, \bibinfo{person}{Junxiao Song}, \bibinfo{person}{Kai Dong}, \bibinfo{person}{Kai Hu}, \bibinfo{person}{Kaige Gao}, \bibinfo{person}{Kang Guan}, \bibinfo{person}{Kexin Huang}, \bibinfo{person}{Kuai Yu}, \bibinfo{person}{Lean Wang}, \bibinfo{person}{Lecong Zhang}, \bibinfo{person}{Lei Xu}, \bibinfo{person}{Leyi Xia}, \bibinfo{person}{Liang Zhao}, \bibinfo{person}{Litong Wang}, \bibinfo{person}{Liyue Zhang}, \bibinfo{person}{Meng Li}, \bibinfo{person}{Miaojun Wang}, \bibinfo{person}{Mingchuan Zhang}, \bibinfo{person}{Minghua Zhang}, \bibinfo{person}{Minghui Tang}, \bibinfo{person}{Mingming Li}, \bibinfo{person}{Ning Tian}, \bibinfo{person}{Panpan Huang}, \bibinfo{person}{Peiyi
  Wang}, \bibinfo{person}{Peng Zhang}, \bibinfo{person}{Qiancheng Wang}, \bibinfo{person}{Qihao Zhu}, \bibinfo{person}{Qinyu Chen}, \bibinfo{person}{Qiushi Du}, \bibinfo{person}{R.~J. Chen}, \bibinfo{person}{R.~L. Jin}, \bibinfo{person}{Ruiqi Ge}, \bibinfo{person}{Ruisong Zhang}, \bibinfo{person}{Ruizhe Pan}, \bibinfo{person}{Runji Wang}, \bibinfo{person}{Runxin Xu}, \bibinfo{person}{Ruoyu Zhang}, \bibinfo{person}{Ruyi Chen}, \bibinfo{person}{S.~S. Li}, \bibinfo{person}{Shanghao Lu}, \bibinfo{person}{Shangyan Zhou}, \bibinfo{person}{Shanhuang Chen}, \bibinfo{person}{Shaoqing Wu}, \bibinfo{person}{Shengfeng Ye}, \bibinfo{person}{Shengfeng Ye}, \bibinfo{person}{Shirong Ma}, \bibinfo{person}{Shiyu Wang}, \bibinfo{person}{Shuang Zhou}, \bibinfo{person}{Shuiping Yu}, \bibinfo{person}{Shunfeng Zhou}, \bibinfo{person}{Shuting Pan}, \bibinfo{person}{T. Wang}, \bibinfo{person}{Tao Yun}, \bibinfo{person}{Tian Pei}, \bibinfo{person}{Tianyu Sun}, \bibinfo{person}{W.~L. Xiao}, \bibinfo{person}{Wangding Zeng},
  \bibinfo{person}{Wanjia Zhao}, \bibinfo{person}{Wei An}, \bibinfo{person}{Wen Liu}, \bibinfo{person}{Wenfeng Liang}, \bibinfo{person}{Wenjun Gao}, \bibinfo{person}{Wenqin Yu}, \bibinfo{person}{Wentao Zhang}, \bibinfo{person}{X.~Q. Li}, \bibinfo{person}{Xiangyue Jin}, \bibinfo{person}{Xianzu Wang}, \bibinfo{person}{Xiao Bi}, \bibinfo{person}{Xiaodong Liu}, \bibinfo{person}{Xiaohan Wang}, \bibinfo{person}{Xiaojin Shen}, \bibinfo{person}{Xiaokang Chen}, \bibinfo{person}{Xiaokang Zhang}, \bibinfo{person}{Xiaosha Chen}, \bibinfo{person}{Xiaotao Nie}, \bibinfo{person}{Xiaowen Sun}, \bibinfo{person}{Xiaoxiang Wang}, \bibinfo{person}{Xin Cheng}, \bibinfo{person}{Xin Liu}, \bibinfo{person}{Xin Xie}, \bibinfo{person}{Xingchao Liu}, \bibinfo{person}{Xingkai Yu}, \bibinfo{person}{Xinnan Song}, \bibinfo{person}{Xinxia Shan}, \bibinfo{person}{Xinyi Zhou}, \bibinfo{person}{Xinyu Yang}, \bibinfo{person}{Xinyuan Li}, \bibinfo{person}{Xuecheng Su}, \bibinfo{person}{Xuheng Lin}, \bibinfo{person}{Y.~K. Li},
  \bibinfo{person}{Y.~Q. Wang}, \bibinfo{person}{Y.~X. Wei}, \bibinfo{person}{Y.~X. Zhu}, \bibinfo{person}{Yang Zhang}, \bibinfo{person}{Yanhong Xu}, \bibinfo{person}{Yanhong Xu}, \bibinfo{person}{Yanping Huang}, \bibinfo{person}{Yao Li}, \bibinfo{person}{Yao Zhao}, \bibinfo{person}{Yaofeng Sun}, \bibinfo{person}{Yaohui Li}, \bibinfo{person}{Yaohui Wang}, \bibinfo{person}{Yi Yu}, \bibinfo{person}{Yi Zheng}, \bibinfo{person}{Yichao Zhang}, \bibinfo{person}{Yifan Shi}, \bibinfo{person}{Yiliang Xiong}, \bibinfo{person}{Ying He}, \bibinfo{person}{Ying Tang}, \bibinfo{person}{Yishi Piao}, \bibinfo{person}{Yisong Wang}, \bibinfo{person}{Yixuan Tan}, \bibinfo{person}{Yiyang Ma}, \bibinfo{person}{Yiyuan Liu}, \bibinfo{person}{Yongqiang Guo}, \bibinfo{person}{Yu Wu}, \bibinfo{person}{Yuan Ou}, \bibinfo{person}{Yuchen Zhu}, \bibinfo{person}{Yuduan Wang}, \bibinfo{person}{Yue Gong}, \bibinfo{person}{Yuheng Zou}, \bibinfo{person}{Yujia He}, \bibinfo{person}{Yukun Zha}, \bibinfo{person}{Yunfan Xiong},
  \bibinfo{person}{Yunxian Ma}, \bibinfo{person}{Yuting Yan}, \bibinfo{person}{Yuxiang Luo}, \bibinfo{person}{Yuxiang You}, \bibinfo{person}{Yuxuan Liu}, \bibinfo{person}{Yuyang Zhou}, \bibinfo{person}{Z.~F. Wu}, \bibinfo{person}{Z.~Z. Ren}, \bibinfo{person}{Zehui Ren}, \bibinfo{person}{Zhangli Sha}, \bibinfo{person}{Zhe Fu}, \bibinfo{person}{Zhean Xu}, \bibinfo{person}{Zhen Huang}, \bibinfo{person}{Zhen Zhang}, \bibinfo{person}{Zhenda Xie}, \bibinfo{person}{Zhengyan Zhang}, \bibinfo{person}{Zhewen Hao}, \bibinfo{person}{Zhibin Gou}, \bibinfo{person}{Zhicheng Ma}, \bibinfo{person}{Zhigang Yan}, \bibinfo{person}{Zhihong Shao}, \bibinfo{person}{Zhipeng Xu}, \bibinfo{person}{Zhiyu Wu}, \bibinfo{person}{Zhongyu Zhang}, \bibinfo{person}{Zhuoshu Li}, \bibinfo{person}{Zihui Gu}, \bibinfo{person}{Zijia Zhu}, \bibinfo{person}{Zijun Liu}, \bibinfo{person}{Zilin Li}, \bibinfo{person}{Ziwei Xie}, \bibinfo{person}{Ziyang Song}, \bibinfo{person}{Ziyi Gao}, {and} \bibinfo{person}{Zizheng Pan}.}
  \bibinfo{year}{2025}\natexlab{}.
\newblock \bibinfo{title}{DeepSeek-V3 Technical Report}.
\newblock
\showeprint[arxiv]{2412.19437}~[cs.CL]
\urldef\tempurl%
\url{https://arxiv.org/abs/2412.19437}
\showURL{%
\tempurl}


\bibitem[Frisella et~al\mbox{.}(2025)]%
        {piper}
\bibfield{author}{\bibinfo{person}{Megan Frisella}, \bibinfo{person}{Arvin Oentoro}, \bibinfo{person}{Xiangyu Gao}, \bibinfo{person}{Gilbert Bernstein}, {and} \bibinfo{person}{Stephanie Wang}.} \bibinfo{year}{2025}\natexlab{}.
\newblock \showarticletitle{Piper: Towards Flexible Pipeline Parallelism for PyTorch}. In \bibinfo{booktitle}{\emph{Proceedings of the 4th Workshop on Practical Adoption Challenges of ML for Systems}} (Seoul, Republic of Korea) \emph{(\bibinfo{series}{PACMI '25})}. \bibinfo{publisher}{Association for Computing Machinery}, \bibinfo{address}{New York, NY, USA}, \bibinfo{pages}{1–6}.
\newblock
\showISBNx{9798400722059}
\href{https://doi.org/10.1145/3766882.3767187}{doi:\nolinkurl{10.1145/3766882.3767187}}


\bibitem[Gupta et~al\mbox{.}({[n.\,d.]})]%
        {guptaautosp}
\bibfield{author}{\bibinfo{person}{Ahan Gupta}, \bibinfo{person}{Zhihao Wang}, \bibinfo{person}{Neel Dani}, \bibinfo{person}{Masahiro Tanaka}, \bibinfo{person}{Olatunji Ruwase}, {and} \bibinfo{person}{Minjia Zhang}.} \bibinfo{year}{[n.\,d.]}\natexlab{}.
\newblock \showarticletitle{AutoSP: Unlocking Long-Context LLM Training Via Compiler-Based Sequence Parallelism}. In \bibinfo{booktitle}{\emph{The Fourteenth International Conference on Learning Representations}}.
\newblock


\bibitem[Harlap et~al\mbox{.}(2018)]%
        {1f1b-msr}
\bibfield{author}{\bibinfo{person}{Aaron Harlap}, \bibinfo{person}{Deepak Narayanan}, \bibinfo{person}{Amar Phanishayee}, \bibinfo{person}{Vivek Seshadri}, \bibinfo{person}{Nikhil Devanur}, \bibinfo{person}{Greg Ganger}, {and} \bibinfo{person}{Phil Gibbons}.} \bibinfo{year}{2018}\natexlab{}.
\newblock \bibinfo{title}{PipeDream: Fast and Efficient Pipeline Parallel DNN Training}.
\newblock
\showeprint[arxiv]{1806.03377}~[cs.DC]
\urldef\tempurl%
\url{https://arxiv.org/abs/1806.03377}
\showURL{%
\tempurl}


\bibitem[Hu et~al\mbox{.}(2026)]%
        {tessera-osdi26}
\bibfield{author}{\bibinfo{person}{Weifang Hu}, \bibinfo{person}{Langshi Chen}, \bibinfo{person}{Man Yuan}, \bibinfo{person}{Youyang Yao}, \bibinfo{person}{Xiulong Yuan}, \bibinfo{person}{Li Tian}, \bibinfo{person}{Yong Li}, \bibinfo{person}{Wei Lin}, \bibinfo{person}{Xuanhua Shi}, \bibinfo{person}{Zhengping Qian}, {and} \bibinfo{person}{Jingren Zhou}.} \bibinfo{year}{2026}\natexlab{}.
\newblock \showarticletitle{Tessera: A Holistic Pipeline Parallelism Framework for Trillion-Parameter Heterogeneous MoE Training}. In \bibinfo{booktitle}{\emph{Proceedings of the 20th USENIX Symposium on Operating Systems Design and Implementation (OSDI '26)}}.
\newblock
\newblock
\shownote{To appear}.


\bibitem[Huang et~al\mbox{.}(2019)]%
        {gpipe}
\bibfield{author}{\bibinfo{person}{Yanping Huang}, \bibinfo{person}{Youlong Cheng}, \bibinfo{person}{Ankur Bapna}, \bibinfo{person}{Orhan Firat}, \bibinfo{person}{Mia~Xu Chen}, \bibinfo{person}{Dehao Chen}, \bibinfo{person}{HyoukJoong Lee}, \bibinfo{person}{Jiquan Ngiam}, \bibinfo{person}{Quoc~V. Le}, \bibinfo{person}{Yonghui Wu}, {and} \bibinfo{person}{Zhifeng Chen}.} \bibinfo{year}{2019}\natexlab{}.
\newblock \bibinfo{booktitle}{\emph{GPipe: efficient training of giant neural networks using pipeline parallelism}}.
\newblock \bibinfo{publisher}{Curran Associates Inc.}, \bibinfo{address}{Red Hook, NY, USA}.
\newblock


\bibitem[Jangda et~al\mbox{.}(2022)]%
        {jangda2022breaking}
\bibfield{author}{\bibinfo{person}{Abhinav Jangda}, \bibinfo{person}{Jun Huang}, \bibinfo{person}{Guodong Liu}, \bibinfo{person}{Amir Hossein~Nodehi Sabet}, \bibinfo{person}{Saeed Maleki}, \bibinfo{person}{Youshan Miao}, \bibinfo{person}{Madanlal Musuvathi}, \bibinfo{person}{Todd Mytkowicz}, {and} \bibinfo{person}{Olli Saarikivi}.} \bibinfo{year}{2022}\natexlab{}.
\newblock \showarticletitle{Breaking the computation and communication abstraction barrier in distributed machine learning workloads}. In \bibinfo{booktitle}{\emph{Proceedings of the 27th ACM International Conference on Architectural Support for Programming Languages and Operating Systems}}. \bibinfo{pages}{402--416}.
\newblock


\bibitem[Jia et~al\mbox{.}(2019)]%
        {jia2019beyond}
\bibfield{author}{\bibinfo{person}{Zhihao Jia}, \bibinfo{person}{Matei Zaharia}, {and} \bibinfo{person}{Alex Aiken}.} \bibinfo{year}{2019}\natexlab{}.
\newblock \showarticletitle{Beyond data and model parallelism for deep neural networks.}
\newblock \bibinfo{journal}{\emph{Proceedings of Machine Learning and Systems}}  \bibinfo{volume}{1} (\bibinfo{year}{2019}), \bibinfo{pages}{1--13}.
\newblock


\bibitem[Lai et~al\mbox{.}(2023)]%
        {lai2023merak}
\bibfield{author}{\bibinfo{person}{Zhiquan Lai}, \bibinfo{person}{Shengwei Li}, \bibinfo{person}{Xudong Tang}, \bibinfo{person}{Keshi Ge}, \bibinfo{person}{Weijie Liu}, \bibinfo{person}{Yabo Duan}, \bibinfo{person}{Linbo Qiao}, {and} \bibinfo{person}{Dongsheng Li}.} \bibinfo{year}{2023}\natexlab{}.
\newblock \showarticletitle{Merak: An efficient distributed dnn training framework with automated 3d parallelism for giant foundation models}.
\newblock \bibinfo{journal}{\emph{IEEE Transactions on Parallel and Distributed Systems}} \bibinfo{volume}{34}, \bibinfo{number}{5} (\bibinfo{year}{2023}), \bibinfo{pages}{1466--1478}.
\newblock


\bibitem[Lepikhin et~al\mbox{.}(2020)]%
        {lepikhin2020gshard}
\bibfield{author}{\bibinfo{person}{Dmitry Lepikhin}, \bibinfo{person}{HyoukJoong Lee}, \bibinfo{person}{Yuanzhong Xu}, \bibinfo{person}{Dehao Chen}, \bibinfo{person}{Orhan Firat}, \bibinfo{person}{Yanping Huang}, \bibinfo{person}{Maxim Krikun}, \bibinfo{person}{Noam Shazeer}, {and} \bibinfo{person}{Zhifeng Chen}.} \bibinfo{year}{2020}\natexlab{}.
\newblock \showarticletitle{Gshard: Scaling giant models with conditional computation and automatic sharding}.
\newblock \bibinfo{journal}{\emph{arXiv preprint arXiv:2006.16668}} (\bibinfo{year}{2020}).
\newblock


\bibitem[Li et~al\mbox{.}(2020)]%
        {ddp-torch}
\bibfield{author}{\bibinfo{person}{Shen Li}, \bibinfo{person}{Yanli Zhao}, \bibinfo{person}{Rohan Varma}, \bibinfo{person}{Omkar Salpekar}, \bibinfo{person}{Pieter Noordhuis}, \bibinfo{person}{Teng Li}, \bibinfo{person}{Adam Paszke}, \bibinfo{person}{Jeff Smith}, \bibinfo{person}{Brian Vaughan}, \bibinfo{person}{Pritam Damania}, {and} \bibinfo{person}{Soumith Chintala}.} \bibinfo{year}{2020}\natexlab{}.
\newblock \bibinfo{title}{PyTorch Distributed: Experiences on Accelerating Data Parallel Training}.
\newblock
\showeprint[arxiv]{2006.15704}~[cs.DC]
\urldef\tempurl%
\url{https://arxiv.org/abs/2006.15704}
\showURL{%
\tempurl}


\bibitem[Liang et~al\mbox{.}(2025)]%
        {liang2025torchtitan}
\bibfield{author}{\bibinfo{person}{Wanchao Liang}, \bibinfo{person}{Tianyu Liu}, \bibinfo{person}{Less Wright}, \bibinfo{person}{Will Constable}, \bibinfo{person}{Andrew Gu}, \bibinfo{person}{Chien-Chin Huang}, \bibinfo{person}{Iris Zhang}, \bibinfo{person}{Wei Feng}, \bibinfo{person}{Howard Huang}, \bibinfo{person}{Junjie Wang}, \bibinfo{person}{Sanket Purandare}, \bibinfo{person}{Gokul Nadathur}, {and} \bibinfo{person}{Stratos Idreos}.} \bibinfo{year}{2025}\natexlab{}.
\newblock \showarticletitle{TorchTitan: One-stop PyTorch native solution for production ready {LLM} pretraining}. In \bibinfo{booktitle}{\emph{The Thirteenth International Conference on Learning Representations}}.
\newblock
\urldef\tempurl%
\url{https://openreview.net/forum?id=SFN6Wm7YBI}
\showURL{%
\tempurl}


\bibitem[Lightning.ai({[n.\,d.]})]%
        {pytorch-stalls}
\bibfield{author}{\bibinfo{person}{Lightning.ai}.} \bibinfo{year}{[n.\,d.]}\natexlab{}.
\newblock \bibinfo{title}{{F}aster {P}y{T}orch {T}raining by {R}educing {P}eak {M}emory (combining backward pass + optimizer step) - {L}ightning {A}{I} --- lightning.ai}.
\newblock \bibinfo{howpublished}{\url{https://lightning.ai/pages/community/tutorial/faster-pytorch-training-by-reducing-peak-memory/}}.
\newblock
\newblock
\shownote{[Accessed 23-04-2026]}.


\bibitem[Lin et~al\mbox{.}(2024)]%
        {lin2024nnscaler}
\bibfield{author}{\bibinfo{person}{Zhiqi Lin}, \bibinfo{person}{Youshan Miao}, \bibinfo{person}{Quanlu Zhang}, \bibinfo{person}{Fan Yang}, \bibinfo{person}{Yi Zhu}, \bibinfo{person}{Cheng Li}, \bibinfo{person}{Saeed Maleki}, \bibinfo{person}{Xu Cao}, \bibinfo{person}{Ning Shang}, \bibinfo{person}{Yilei Yang}, {et~al\mbox{.}}} \bibinfo{year}{2024}\natexlab{}.
\newblock \showarticletitle{$\{$nnScaler$\}$:$\{$Constraint-Guided$\}$ Parallelization Plan Generation for Deep Learning Training}. In \bibinfo{booktitle}{\emph{18th USENIX Symposium on Operating Systems Design and Implementation (OSDI 24)}}. \bibinfo{pages}{347--363}.
\newblock


\bibitem[Liu et~al\mbox{.}(2024)]%
        {liu2024aceso}
\bibfield{author}{\bibinfo{person}{Guodong Liu}, \bibinfo{person}{Youshan Miao}, \bibinfo{person}{Zhiqi Lin}, \bibinfo{person}{Xiaoxiang Shi}, \bibinfo{person}{Saeed Maleki}, \bibinfo{person}{Fan Yang}, \bibinfo{person}{Yungang Bao}, {and} \bibinfo{person}{Sa Wang}.} \bibinfo{year}{2024}\natexlab{}.
\newblock \showarticletitle{Aceso: Efficient parallel DNN training through iterative bottleneck alleviation}. In \bibinfo{booktitle}{\emph{Proceedings of the Nineteenth European Conference on Computer Systems}}. \bibinfo{pages}{163--181}.
\newblock


\bibitem[Liu et~al\mbox{.}(2023)]%
        {llava-model}
\bibfield{author}{\bibinfo{person}{Haotian Liu}, \bibinfo{person}{Chunyuan Li}, \bibinfo{person}{Qingyang Wu}, {and} \bibinfo{person}{Yong~Jae Lee}.} \bibinfo{year}{2023}\natexlab{}.
\newblock \showarticletitle{Visual Instruction Tuning}. In \bibinfo{booktitle}{\emph{Advances in Neural Information Processing Systems}}, \bibfield{editor}{\bibinfo{person}{A.~Oh}, \bibinfo{person}{T.~Naumann}, \bibinfo{person}{A.~Globerson}, \bibinfo{person}{K.~Saenko}, \bibinfo{person}{M.~Hardt}, {and} \bibinfo{person}{S.~Levine}} (Eds.), Vol.~\bibinfo{volume}{36}. \bibinfo{publisher}{Curran Associates, Inc.}, \bibinfo{pages}{34892--34916}.
\newblock
\urldef\tempurl%
\url{https://proceedings.neurips.cc/paper_files/paper/2023/file/6dcf277ea32ce3288914faf369fe6de0-Paper-Conference.pdf}
\showURL{%
\tempurl}


\bibitem[Miao et~al\mbox{.}(2022)]%
        {miao2022galvatron}
\bibfield{author}{\bibinfo{person}{Xupeng Miao}, \bibinfo{person}{Yujie Wang}, \bibinfo{person}{Youhe Jiang}, \bibinfo{person}{Chunan Shi}, \bibinfo{person}{Xiaonan Nie}, \bibinfo{person}{Hailin Zhang}, {and} \bibinfo{person}{Bin Cui}.} \bibinfo{year}{2022}\natexlab{}.
\newblock \showarticletitle{Galvatron: Efficient transformer training over multiple gpus using automatic parallelism}.
\newblock \bibinfo{journal}{\emph{arXiv preprint arXiv:2211.13878}} (\bibinfo{year}{2022}).
\newblock


\bibitem[Narayanan et~al\mbox{.}(2021a)]%
        {pipedream}
\bibfield{author}{\bibinfo{person}{Deepak Narayanan}, \bibinfo{person}{Amar Phanishayee}, \bibinfo{person}{Kaiyu Shi}, \bibinfo{person}{Xie Chen}, {and} \bibinfo{person}{Matei Zaharia}.} \bibinfo{year}{2021}\natexlab{a}.
\newblock \showarticletitle{Memory-Efficient Pipeline-Parallel DNN Training}. In \bibinfo{booktitle}{\emph{Proceedings of the 38th International Conference on Machine Learning}} \emph{(\bibinfo{series}{Proceedings of Machine Learning Research}, Vol.~\bibinfo{volume}{139})}, \bibfield{editor}{\bibinfo{person}{Marina Meila} {and} \bibinfo{person}{Tong Zhang}} (Eds.). \bibinfo{publisher}{PMLR}, \bibinfo{pages}{7937--7947}.
\newblock
\urldef\tempurl%
\url{https://proceedings.mlr.press/v139/narayanan21a.html}
\showURL{%
\tempurl}


\bibitem[Narayanan et~al\mbox{.}(2021b)]%
        {i1f1b-nvidia}
\bibfield{author}{\bibinfo{person}{Deepak Narayanan}, \bibinfo{person}{Mohammad Shoeybi}, \bibinfo{person}{Jared Casper}, \bibinfo{person}{Patrick LeGresley}, \bibinfo{person}{Mostofa Patwary}, \bibinfo{person}{Vijay~Anand Korthikanti}, \bibinfo{person}{Dmitri Vainbrand}, \bibinfo{person}{Prethvi Kashinkunti}, \bibinfo{person}{Julie Bernauer}, \bibinfo{person}{Bryan Catanzaro}, \bibinfo{person}{Amar Phanishayee}, {and} \bibinfo{person}{Matei Zaharia}.} \bibinfo{year}{2021}\natexlab{b}.
\newblock \bibinfo{title}{Efficient Large-Scale Language Model Training on GPU Clusters Using Megatron-LM}.
\newblock
\showeprint[arxiv]{2104.04473}~[cs.CL]
\urldef\tempurl%
\url{https://arxiv.org/abs/2104.04473}
\showURL{%
\tempurl}


\bibitem[NVIDIA({[n.\,d.]})]%
        {nccl}
\bibfield{author}{\bibinfo{person}{NVIDIA}.} \bibinfo{year}{[n.\,d.]}\natexlab{}.
\newblock \bibinfo{title}{NVIDIA Collective Communications Library (NCCL)}.
\newblock \bibinfo{howpublished}{\url{https://developer.nvidia.com/nccl}}.
\newblock


\bibitem[Pan et~al\mbox{.}(2026)]%
        {dynaflow}
\bibfield{author}{\bibinfo{person}{Yi Pan}, \bibinfo{person}{Yile Gu}, \bibinfo{person}{Jinbin Luo}, \bibinfo{person}{Yibo Wu}, \bibinfo{person}{Ziren Wang}, \bibinfo{person}{Hongtao Zhang}, \bibinfo{person}{Ziyi Xu}, \bibinfo{person}{Shengkai Lin}, \bibinfo{person}{Baris Kasikci}, {and} \bibinfo{person}{Stephanie Wang}.} \bibinfo{year}{2026}\natexlab{}.
\newblock \showarticletitle{DynaFlow: Transparent and Flexible Intra-Device Parallelism via Programmable Operator Scheduling}.
\newblock \bibinfo{journal}{\emph{Proceedings of Machine Learning and Systems}}.
\newblock
\newblock
\shownote{To appear}.


\bibitem[Peng et~al\mbox{.}(2019)]%
        {peng2019generic}
\bibfield{author}{\bibinfo{person}{Yanghua Peng}, \bibinfo{person}{Yibo Zhu}, \bibinfo{person}{Yangrui Chen}, \bibinfo{person}{Yixin Bao}, \bibinfo{person}{Bairen Yi}, \bibinfo{person}{Chang Lan}, \bibinfo{person}{Chuan Wu}, {and} \bibinfo{person}{Chuanxiong Guo}.} \bibinfo{year}{2019}\natexlab{}.
\newblock \showarticletitle{A generic communication scheduler for distributed DNN training acceleration}. In \bibinfo{booktitle}{\emph{Proceedings of the 27th ACM Symposium on Operating Systems Principles}}. \bibinfo{pages}{16--29}.
\newblock


\bibitem[Qi et~al\mbox{.}(2023)]%
        {zerobubble}
\bibfield{author}{\bibinfo{person}{Penghui Qi}, \bibinfo{person}{Xinyi Wan}, \bibinfo{person}{Guangxing Huang}, {and} \bibinfo{person}{Min Lin}.} \bibinfo{year}{2023}\natexlab{}.
\newblock \showarticletitle{Zero Bubble Pipeline Parallelism}.
\newblock \bibinfo{journal}{\emph{ArXiv}}  \bibinfo{volume}{abs/2401.10241} (\bibinfo{year}{2023}).
\newblock
\urldef\tempurl%
\url{https://api.semanticscholar.org/CorpusID:267060979}
\showURL{%
\tempurl}


\bibitem[Qi et~al\mbox{.}(2025)]%
        {qi2025dual}
\bibfield{author}{\bibinfo{person}{Penghui Qi}, \bibinfo{person}{Xinyi Wan}, \bibinfo{person}{Guangxing Huang}, {and} \bibinfo{person}{Min Lin}.} \bibinfo{year}{2025}\natexlab{}.
\newblock \bibinfo{title}{DualPipe could be better without the Dual}.
\newblock \bibinfo{howpublished}{\url{https://hackmd.io/@ufotalent/r1lVXsa9Jg}}.
\newblock
\newblock
\shownote{Blog}.


\bibitem[Radford et~al\mbox{.}(2021)]%
        {clip-model}
\bibfield{author}{\bibinfo{person}{Alec Radford}, \bibinfo{person}{Jong~Wook Kim}, \bibinfo{person}{Chris Hallacy}, \bibinfo{person}{Aditya Ramesh}, \bibinfo{person}{Gabriel Goh}, \bibinfo{person}{Sandhini Agarwal}, \bibinfo{person}{Girish Sastry}, \bibinfo{person}{Amanda Askell}, \bibinfo{person}{Pamela Mishkin}, \bibinfo{person}{Jack Clark}, \bibinfo{person}{Gretchen Krueger}, {and} \bibinfo{person}{Ilya Sutskever}.} \bibinfo{year}{2021}\natexlab{}.
\newblock \showarticletitle{Learning Transferable Visual Models From Natural Language Supervision}. In \bibinfo{booktitle}{\emph{Proceedings of the 38th International Conference on Machine Learning}} \emph{(\bibinfo{series}{Proceedings of Machine Learning Research}, Vol.~\bibinfo{volume}{139})}, \bibfield{editor}{\bibinfo{person}{Marina Meila} {and} \bibinfo{person}{Tong Zhang}} (Eds.). \bibinfo{publisher}{PMLR}, \bibinfo{pages}{8748--8763}.
\newblock
\urldef\tempurl%
\url{https://proceedings.mlr.press/v139/radford21a.html}
\showURL{%
\tempurl}


\bibitem[Ragan-Kelley et~al\mbox{.}(2013)]%
        {halide}
\bibfield{author}{\bibinfo{person}{Jonathan Ragan-Kelley}, \bibinfo{person}{Connelly Barnes}, \bibinfo{person}{Andrew Adams}, \bibinfo{person}{Sylvain Paris}, \bibinfo{person}{Fr\'{e}do Durand}, {and} \bibinfo{person}{Saman Amarasinghe}.} \bibinfo{year}{2013}\natexlab{}.
\newblock \showarticletitle{Halide: a language and compiler for optimizing parallelism, locality, and recomputation in image processing pipelines}.
\newblock \bibinfo{journal}{\emph{SIGPLAN Not.}} \bibinfo{volume}{48}, \bibinfo{number}{6} (\bibinfo{date}{June} \bibinfo{year}{2013}), \bibinfo{pages}{519–530}.
\newblock
\showISSN{0362-1340}
\href{https://doi.org/10.1145/2499370.2462176}{doi:\nolinkurl{10.1145/2499370.2462176}}


\bibitem[Rajbhandari et~al\mbox{.}(2019)]%
        {zero-deepspeed}
\bibfield{author}{\bibinfo{person}{Samyam Rajbhandari}, \bibinfo{person}{Jeff Rasley}, \bibinfo{person}{Olatunji Ruwase}, {and} \bibinfo{person}{Yuxiong He}.} \bibinfo{year}{2019}\natexlab{}.
\newblock \showarticletitle{ZeRO: Memory Optimization Towards Training {A} Trillion Parameter Models}.
\newblock \bibinfo{journal}{\emph{CoRR}}  \bibinfo{volume}{abs/1910.02054} (\bibinfo{year}{2019}).
\newblock
\showeprint[arXiv]{1910.02054}
\urldef\tempurl%
\url{http://arxiv.org/abs/1910.02054}
\showURL{%
\tempurl}


\bibitem[Reed et~al\mbox{.}(2022)]%
        {pippy2022}
\bibfield{author}{\bibinfo{person}{James Reed}, \bibinfo{person}{Pavel Belevich}, \bibinfo{person}{Ke Wen}, \bibinfo{person}{Howard Huang}, {and} \bibinfo{person}{Will Constable}.} \bibinfo{year}{2022}\natexlab{}.
\newblock \bibinfo{title}{PiPPy: Pipeline Parallelism for PyTorch}.
\newblock \bibinfo{howpublished}{\url{https://github.com/pytorch/PiPPy}}.
\newblock


\bibitem[Shoeybi et~al\mbox{.}(2019)]%
        {megatron-nvidia}
\bibfield{author}{\bibinfo{person}{Mohammad Shoeybi}, \bibinfo{person}{Mostofa Patwary}, \bibinfo{person}{Raul Puri}, \bibinfo{person}{Patrick LeGresley}, \bibinfo{person}{Jared Casper}, {and} \bibinfo{person}{Bryan Catanzaro}.} \bibinfo{year}{2019}\natexlab{}.
\newblock \showarticletitle{Megatron-LM: Training Multi-Billion Parameter Language Models Using Model Parallelism}.
\newblock \bibinfo{journal}{\emph{CoRR}}  \bibinfo{volume}{abs/1909.08053} (\bibinfo{year}{2019}).
\newblock
\showeprint[arXiv]{1909.08053}
\urldef\tempurl%
\url{http://arxiv.org/abs/1909.08053}
\showURL{%
\tempurl}


\bibitem[Sun et~al\mbox{.}(2024)]%
        {sun2024adapipe}
\bibfield{author}{\bibinfo{person}{Zhenbo Sun}, \bibinfo{person}{Huanqi Cao}, \bibinfo{person}{Yuanwei Wang}, \bibinfo{person}{Guanyu Feng}, \bibinfo{person}{Shengqi Chen}, \bibinfo{person}{Haojie Wang}, {and} \bibinfo{person}{Wenguang Chen}.} \bibinfo{year}{2024}\natexlab{}.
\newblock \showarticletitle{Adapipe: Optimizing pipeline parallelism with adaptive recomputation and partitioning}. In \bibinfo{booktitle}{\emph{Proceedings of the 29th ACM International Conference on Architectural Support for Programming Languages and Operating Systems, Volume 3}}. \bibinfo{pages}{86--100}.
\newblock


\bibitem[Tanaka et~al\mbox{.}(2025)]%
        {tanaka2025deepcompile}
\bibfield{author}{\bibinfo{person}{Masahiro Tanaka}, \bibinfo{person}{Du Li}, \bibinfo{person}{Umesh Chand}, \bibinfo{person}{Ali Zafar}, \bibinfo{person}{Haiying Shen}, {and} \bibinfo{person}{Olatunji Ruwase}.} \bibinfo{year}{2025}\natexlab{}.
\newblock \showarticletitle{DeepCompile: A Compiler-Driven Approach to Optimizing Distributed Deep Learning Training}.
\newblock \bibinfo{journal}{\emph{arXiv preprint arXiv:2504.09983}} (\bibinfo{year}{2025}).
\newblock


\bibitem[Tarnawski et~al\mbox{.}(2021)]%
        {tarnawski2021piper}
\bibfield{author}{\bibinfo{person}{Jakub~M Tarnawski}, \bibinfo{person}{Deepak Narayanan}, {and} \bibinfo{person}{Amar Phanishayee}.} \bibinfo{year}{2021}\natexlab{}.
\newblock \showarticletitle{Piper: Multidimensional planner for dnn parallelization}.
\newblock \bibinfo{journal}{\emph{Advances in Neural Information Processing Systems}}  \bibinfo{volume}{34} (\bibinfo{year}{2021}), \bibinfo{pages}{24829--24840}.
\newblock


\bibitem[Unger et~al\mbox{.}(2022)]%
        {unger2022unity}
\bibfield{author}{\bibinfo{person}{Colin Unger}, \bibinfo{person}{Zhihao Jia}, \bibinfo{person}{Wei Wu}, \bibinfo{person}{Sina Lin}, \bibinfo{person}{Mandeep Baines}, \bibinfo{person}{Carlos Efrain~Quintero Narvaez}, \bibinfo{person}{Vinay Ramakrishnaiah}, \bibinfo{person}{Nirmal Prajapati}, \bibinfo{person}{Pat McCormick}, \bibinfo{person}{Jamaludin Mohd-Yusof}, {et~al\mbox{.}}} \bibinfo{year}{2022}\natexlab{}.
\newblock \showarticletitle{Unity: Accelerating $\{$DNN$\}$ training through joint optimization of algebraic transformations and parallelization}. In \bibinfo{booktitle}{\emph{16th USENIX Symposium on Operating Systems Design and Implementation (OSDI 22)}}. \bibinfo{pages}{267--284}.
\newblock


\bibitem[Wang et~al\mbox{.}(2019)]%
        {wang2019supporting}
\bibfield{author}{\bibinfo{person}{Minjie Wang}, \bibinfo{person}{Chien-chin Huang}, {and} \bibinfo{person}{Jinyang Li}.} \bibinfo{year}{2019}\natexlab{}.
\newblock \showarticletitle{Supporting very large models using automatic dataflow graph partitioning}. In \bibinfo{booktitle}{\emph{Proceedings of the Fourteenth EuroSys Conference 2019}}. \bibinfo{pages}{1--17}.
\newblock


\bibitem[Wang et~al\mbox{.}(2022)]%
        {wang2022overlap}
\bibfield{author}{\bibinfo{person}{Shibo Wang}, \bibinfo{person}{Jinliang Wei}, \bibinfo{person}{Amit Sabne}, \bibinfo{person}{Andy Davis}, \bibinfo{person}{Berkin Ilbeyi}, \bibinfo{person}{Blake Hechtman}, \bibinfo{person}{Dehao Chen}, \bibinfo{person}{Karthik~Srinivasa Murthy}, \bibinfo{person}{Marcello Maggioni}, \bibinfo{person}{Qiao Zhang}, {et~al\mbox{.}}} \bibinfo{year}{2022}\natexlab{}.
\newblock \showarticletitle{Overlap communication with dependent computation via decomposition in large deep learning models}. In \bibinfo{booktitle}{\emph{Proceedings of the 28th ACM International Conference on Architectural Support for Programming Languages and Operating Systems, Volume 1}}. \bibinfo{pages}{93--106}.
\newblock


\bibitem[Wang et~al\mbox{.}(2024)]%
        {asynctp}
\bibfield{author}{\bibinfo{person}{Yifu Wang}, \bibinfo{person}{Horace He}, \bibinfo{person}{Less Wright}, \bibinfo{person}{Luca Wehrstedt}, \bibinfo{person}{Tianyu Liu}, {and} \bibinfo{person}{Wanchao Liang}.} \bibinfo{year}{2024}\natexlab{}.
\newblock \bibinfo{title}{Distributed w/ TorchTitan: Introducing Async Tensor Parallelism in PyTorch}.
\newblock \bibinfo{howpublished}{\url{https://discuss.pytorch.org/t/distributed-w-torchtitan-introducing-async-tensor-parallelism-in-pytorch/209487}}.
\newblock
\newblock
\shownote{PyTorch Forum Post. Accessed: 2026-04-23}.


\bibitem[Xhebraj et~al\mbox{.}(2025)]%
        {jaxpp-nvidia}
\bibfield{author}{\bibinfo{person}{Anxhelo Xhebraj}, \bibinfo{person}{Sean Lee}, \bibinfo{person}{Hanfeng Chen}, {and} \bibinfo{person}{Vinod Grover}.} \bibinfo{year}{2025}\natexlab{}.
\newblock \showarticletitle{Scaling deep learning training with MPMD pipeline parallelism}.
\newblock \bibinfo{journal}{\emph{Proceedings of Machine Learning and Systems}}  \bibinfo{volume}{7} (\bibinfo{year}{2025}).
\newblock


\bibitem[Xu et~al\mbox{.}(2021)]%
        {gspmd-google}
\bibfield{author}{\bibinfo{person}{Yuanzhong Xu}, \bibinfo{person}{HyoukJoong Lee}, \bibinfo{person}{Dehao Chen}, \bibinfo{person}{Blake Hechtman}, \bibinfo{person}{Yanping Huang}, \bibinfo{person}{Rahul Joshi}, \bibinfo{person}{Maxim Krikun}, \bibinfo{person}{Dmitry Lepikhin}, \bibinfo{person}{Andy Ly}, \bibinfo{person}{Marcello Maggioni}, \bibinfo{person}{Ruoming Pang}, \bibinfo{person}{Noam Shazeer}, \bibinfo{person}{Shibo Wang}, \bibinfo{person}{Tao Wang}, \bibinfo{person}{Yonghui Wu}, {and} \bibinfo{person}{Zhifeng Chen}.} \bibinfo{year}{2021}\natexlab{}.
\newblock \bibinfo{title}{GSPMD: General and Scalable Parallelization for ML Computation Graphs}.
\newblock
\showeprint[arxiv]{2105.04663}~[cs.DC]
\urldef\tempurl%
\url{https://arxiv.org/abs/2105.04663}
\showURL{%
\tempurl}


\bibitem[Yang et~al\mbox{.}(2025)]%
        {context_parallelism}
\bibfield{author}{\bibinfo{person}{Amy Yang}, \bibinfo{person}{Jingyi Yang}, \bibinfo{person}{Aya Ibrahim}, \bibinfo{person}{Xinfeng Xie}, \bibinfo{person}{Bangsheng Tang}, \bibinfo{person}{Grigory Sizov}, \bibinfo{person}{Jongsoo Park}, {and} \bibinfo{person}{Jianyu Huang}.} \bibinfo{year}{2025}\natexlab{}.
\newblock \showarticletitle{Context Parallelism for Scalable Million-Token Inference}. In \bibinfo{booktitle}{\emph{Proceedings of Machine Learning and Systems}}, \bibfield{editor}{\bibinfo{person}{M.~Zaharia}, \bibinfo{person}{G.~Joshi}, {and} \bibinfo{person}{Y.~Lin}} (Eds.), Vol.~\bibinfo{volume}{7}. \bibinfo{publisher}{MLSys}.
\newblock
\urldef\tempurl%
\url{https://proceedings.mlsys.org/paper_files/paper/2025/file/78834433edc3291f4c6cbbd2759324db-Paper-Conference.pdf}
\showURL{%
\tempurl}


\bibitem[Yuan et~al\mbox{.}(2024)]%
        {yuan2024accelerating}
\bibfield{author}{\bibinfo{person}{Tailing Yuan}, \bibinfo{person}{Yuliang Liu}, \bibinfo{person}{Xucheng Ye}, \bibinfo{person}{Shenglong Zhang}, \bibinfo{person}{Jianchao Tan}, \bibinfo{person}{Bin Chen}, \bibinfo{person}{Chengru Song}, {and} \bibinfo{person}{Di Zhang}.} \bibinfo{year}{2024}\natexlab{}.
\newblock \showarticletitle{Accelerating the training of large language models using efficient activation rematerialization and optimal hybrid parallelism}. In \bibinfo{booktitle}{\emph{2024 USENIX Annual Technical Conference (USENIX ATC 24)}}. \bibinfo{pages}{545--561}.
\newblock


\bibitem[Zhang et~al\mbox{.}(2024)]%
        {mmllms}
\bibfield{author}{\bibinfo{person}{Duzhen Zhang}, \bibinfo{person}{Yahan Yu}, \bibinfo{person}{Jiahua Dong}, \bibinfo{person}{Chenxing Li}, \bibinfo{person}{Dan Su}, \bibinfo{person}{Chenhui Chu}, {and} \bibinfo{person}{Dong Yu}.} \bibinfo{year}{2024}\natexlab{}.
\newblock \showarticletitle{{MM}-{LLM}s: Recent Advances in {M}ulti{M}odal Large Language Models}. In \bibinfo{booktitle}{\emph{Findings of the Association for Computational Linguistics: ACL 2024}}, \bibfield{editor}{\bibinfo{person}{Lun-Wei Ku}, \bibinfo{person}{Andre Martins}, {and} \bibinfo{person}{Vivek Srikumar}} (Eds.). \bibinfo{publisher}{Association for Computational Linguistics}, \bibinfo{address}{Bangkok, Thailand}, \bibinfo{pages}{12401--12430}.
\newblock
\href{https://doi.org/10.18653/v1/2024.findings-acl.738}{doi:\nolinkurl{10.18653/v1/2024.findings-acl.738}}


\bibitem[Zhang et~al\mbox{.}(2025)]%
        {disttrain}
\bibfield{author}{\bibinfo{person}{Zili Zhang}, \bibinfo{person}{Yinmin Zhong}, \bibinfo{person}{Yimin Jiang}, \bibinfo{person}{Hanpeng Hu}, \bibinfo{person}{Jianjian Sun}, \bibinfo{person}{Zheng Ge}, \bibinfo{person}{Yibo Zhu}, \bibinfo{person}{Daxin Jiang}, {and} \bibinfo{person}{Xin Jin}.} \bibinfo{year}{2025}\natexlab{}.
\newblock \showarticletitle{DistTrain: Addressing Model and Data Heterogeneity with Disaggregated Training for Multimodal Large Language Models}. In \bibinfo{booktitle}{\emph{Proceedings of the ACM SIGCOMM 2025 Conference}} (S\~{a}o Francisco Convent, Coimbra, Portugal) \emph{(\bibinfo{series}{SIGCOMM '25})}. \bibinfo{publisher}{Association for Computing Machinery}, \bibinfo{address}{New York, NY, USA}, \bibinfo{pages}{24–38}.
\newblock
\showISBNx{9798400715242}
\href{https://doi.org/10.1145/3718958.3750472}{doi:\nolinkurl{10.1145/3718958.3750472}}


\bibitem[Zhao et~al\mbox{.}(2025)]%
        {deepep2025}
\bibfield{author}{\bibinfo{person}{Chenggang Zhao}, \bibinfo{person}{Shangyan Zhou}, \bibinfo{person}{Liyue Zhang}, \bibinfo{person}{Chengqi Deng}, \bibinfo{person}{Zhean Xu}, \bibinfo{person}{Yuxuan Liu}, \bibinfo{person}{Kuai Yu}, \bibinfo{person}{Jiashi Li}, {and} \bibinfo{person}{Liang Zhao}.} \bibinfo{year}{2025}\natexlab{}.
\newblock \bibinfo{title}{DeepEP: an efficient expert-parallel communication library}.
\newblock \bibinfo{howpublished}{\url{https://github.com/deepseek-ai/DeepEP}}.
\newblock


\bibitem[Zhao et~al\mbox{.}(2023)]%
        {fsdp-pytorch}
\bibfield{author}{\bibinfo{person}{Yanli Zhao}, \bibinfo{person}{Andrew Gu}, \bibinfo{person}{Rohan Varma}, \bibinfo{person}{Liang Luo}, \bibinfo{person}{Chien-Chin Huang}, \bibinfo{person}{Min Xu}, \bibinfo{person}{Less Wright}, \bibinfo{person}{Hamid Shojanazeri}, \bibinfo{person}{Myle Ott}, \bibinfo{person}{Sam Shleifer}, \bibinfo{person}{Alban Desmaison}, \bibinfo{person}{Can Balioglu}, \bibinfo{person}{Pritam Damania}, \bibinfo{person}{Bernard Nguyen}, \bibinfo{person}{Geeta Chauhan}, \bibinfo{person}{Yuchen Hao}, \bibinfo{person}{Ajit Mathews}, {and} \bibinfo{person}{Shen Li}.} \bibinfo{year}{2023}\natexlab{}.
\newblock \bibinfo{title}{PyTorch FSDP: Experiences on Scaling Fully Sharded Data Parallel}.
\newblock
\showeprint[arxiv]{2304.11277}~[cs.DC]
\urldef\tempurl%
\url{https://arxiv.org/abs/2304.11277}
\showURL{%
\tempurl}


\bibitem[Zheng et~al\mbox{.}(2022)]%
        {alpa-ucb}
\bibfield{author}{\bibinfo{person}{Lianmin Zheng}, \bibinfo{person}{Zhuohan Li}, \bibinfo{person}{Hao Zhang}, \bibinfo{person}{Yonghao Zhuang}, \bibinfo{person}{Zhifeng Chen}, \bibinfo{person}{Yanping Huang}, \bibinfo{person}{Yida Wang}, \bibinfo{person}{Yuanzhong Xu}, \bibinfo{person}{Danyang Zhuo}, \bibinfo{person}{Eric~P Xing}, {et~al\mbox{.}}} \bibinfo{year}{2022}\natexlab{}.
\newblock \showarticletitle{Alpa: Automating inter-and $\{$Intra-Operator$\}$ parallelism for distributed deep learning}. In \bibinfo{booktitle}{\emph{16th USENIX Symposium on Operating Systems Design and Implementation (OSDI 22)}}. \bibinfo{pages}{559--578}.
\newblock


\bibitem[Zhu et~al\mbox{.}(2025)]%
        {zhu2025mist}
\bibfield{author}{\bibinfo{person}{Zhanda Zhu}, \bibinfo{person}{Christina Giannoula}, \bibinfo{person}{Muralidhar Andoorveedu}, \bibinfo{person}{Qidong Su}, \bibinfo{person}{Karttikeya Mangalam}, \bibinfo{person}{Bojian Zheng}, {and} \bibinfo{person}{Gennady Pekhimenko}.} \bibinfo{year}{2025}\natexlab{}.
\newblock \showarticletitle{Mist: Efficient Distributed Training of Large Language Models via Memory-Parallelism Co-Optimization}. In \bibinfo{booktitle}{\emph{Proceedings of the Twentieth European Conference on Computer Systems}}. \bibinfo{pages}{1298--1316}.
\newblock


\end{thebibliography}

% \appendix

\end{document}